\documentclass[12pt]{article}
\usepackage{isolatin1}
\usepackage{graphicx}
\usepackage{indentfirst}
\usepackage{setspace}

\begin{document}

\setlength{\hoffset}{-1.8 cm}
\setlength{\voffset}{-3 cm}

\def\etal{\emph{et al }}
\def\ibid{\emph{ibid.}}

\title{Capillary Condensation in Confined Media}

\author{Elisabeth Charlaix$^1$ and Matteo Ciccotti$^{2,*}$}

\date{Handbook of Nanophysics - Volume 1 \\ Edited by Klaus Sattler \\ CRC Press (To appear in June 2010)}

\maketitle

\date{$^1$Laboratoire de Physique de la Matière Condensée et Nanostructures, UMR5586, CNRS, Université Claude Bernard Lyon 1,
Domaine Scientifique de la Doua, Bâtiment Léon Brillouin, 43 Boulevard du 11 novembre 1918, 69622, Villeurbanne, France}

\date{$^2$Laboratoire des Colloïdes, Verres et Nanomatériaux, UMR 5587, CNRS, Université Montpellier 2, Place Bataillon, cc26, 34095, Montpellier, France}

\

\date{Keywords : capillary condensation, confined fluids, wetting, SFA, AFM}

\

$^*$ e-mail: \verb|matteo.ciccotti@univ-montp2.fr|
~~~phone: +33-(0)4-67143529

%\doublespacing

\section*{Abstract}
%200 words

We review here the physics of capillary condensation of liquids in confined media, with a special regard to the application in nanotechnologies. The thermodynamics of capillary condensation and thin film adsorption are first exposed along with all the relevant notions. The focus is then shifted to the modelling of capillary forces, to their measurements techniques (including SFA, AFM and crack tips) and to their influence on AFM imaging techniques as well as on the static and dynamic friction properties of solids (including granular heaps and sliding nanocontacts). A great attention is spent in investigating the delicate role of the surface roughness and all the difficulties involved in the reduction of the probe size to nanometric dimensions. Another major consequence of capillary condensation in nanosystems is the activation of several chemical and corrosive processes that can significantly alter the surface properties, such as dissolution/redeposition of solid materials and stress-corrosion crack propagation.

\newpage

\tableofcontents

\newpage

\section{Physics of capillary condensation}

\subsection{Relevance in nano-systems}
\parindent=20pt

As the size of systems goes down, surface effects become increasingly important. Capillary condensation, which results from the effect of surfaces on the phase diagram of a fluid, is an ubiquitous phenomenon at nanoscale, occurring in all confined geometries, divided media, cracks, or contacts between surfaces (Bowden and Tabor 1950). The very large capillary forces induced by highly curved menisci, have strong effect on the mechanical properties of contacts.
The impact of capillary forces in Micro/Nano Electro-Mechanical Systems (MEMS \& NEMS) is huge and often prevents the function of small scale active systems under ambient condition or causes damage during the fabrication process.

\begin{figure}[!h]
\centering
\includegraphics[height=6 cm]{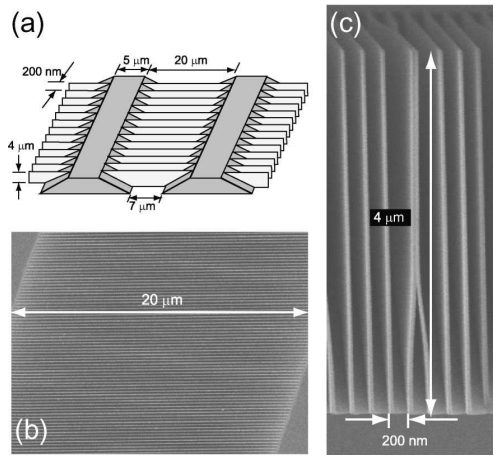}
\caption{Stiction effect due to drying in the nanofbrication of a nanomirror array (c). The spacing between the lamellae is 200 nm as described in (a). If the aspect ratio of the lamellae is larger than a critical value, the stiffness of the lamellae becomes too small to withstand the attractive action of the capillary forces induced by drying at the moment where the liquid meniscus must bend to empty the interlamellar space. We remark that in drying processes the meniscus curvature and capillary pressure are quite smaller than the equilibrium values (cf.\ insert B), but they still have a great impact (Figure after Heilmann \etal 2008).}
\label{fig:Stiction}
\end{figure}

Since the nanocomponents are generally very compliant and present an elevated surface/volume ratio, the capillary forces developing in the confined spaces separating the components when these are exposed to ambient condition can have a dramatic effect in deforming them and preventing their service.
Stiction or adhesion between the substrate (usually silicon based) and the microstructures occurs during the isotropic wet etching of the sacrificial layer  (Bhushan \etal 2003). The capillary forces cause by the surface tension of the liquid between the microstructures (or in the gaps separating them from the substrate) during the drying of the wet etchant cause the two surfaces to adhere together. Fig.\ \ref{fig:Stiction} shows an example of the effect of drying after the nanofabrication of a nanomirror array. Separating the two surfaces is often complicated due to the fragile nature of the microstructures\footnote{Stiction is often circumvented by the use of a sublimating fluid, such as supercritical carbon dioxide.}.

In divided media, capillary forces control the cohesion of the media, but also have dramatic influence on the ageing properties of materials. Since the condensation of liquid bridges is a first order transition, it gives rise to slow activated phenomena that are responsible for long time scale variations of the cohesion forces (cf.\ section \ref{sec:CapillaryForces}). Capillary forces also have a strong effect on the friction properties of sliding nanocontacts where they are responsible for ageing effects and enhanced stick-slip motion (cf.\ section \ref{sec:friction}).
Finally, the presence of capillary menisci and nanometric water films on solid surfaces has deep consequences on the surface physical and chemical properties, notably by permitting the activation of nanoscale corrosion processes, such as local dissolution and recondensation, hydration, oxidation, hydrolysis, and lixiviation. These phenomena can either lead to the long term improvement of the mechanical properties of nanostructured materials by recrystallization of solid joints or to the failure of microsctructures due to stress-corrosion crack propagation (cf.\ section \ref{sec:SurfaceChemistry}).

\subsection {Physics of capillary condensation}
\parindent=20pt

Let us consider  two parallel solid surfaces separated by a distance $D$, in contact with a reservoir of  vapor at a
pressure $P_v$ and temperature $T$. If $D$ is very large, the liquid-vapor equilibrium occurs at  the saturating  pressure $P_v=P_{sat}$.
For a finite $D$, if the surface tension $\gamma_{sl}$   of the wet solid surface (see insert A) is lower than the one $\gamma_{sv}$ of the dry solid surface, the solid favors liquid condensation. One should therefore ask if the solid can successfully stabilize a liquid phase when the vapor phase is stable in the bulk, i.e. $P_{v}<P_{sat}$.
To answer this question, one must compare the
grand canonical potential (see insert A) of two configurations: the `liquid-filled interstice', which
we shall call the condensed state, and the `vapor-filled interstice', i.e.\ the non-condensed state:

\begin{figure}[!h]
\includegraphics[width=15cm]{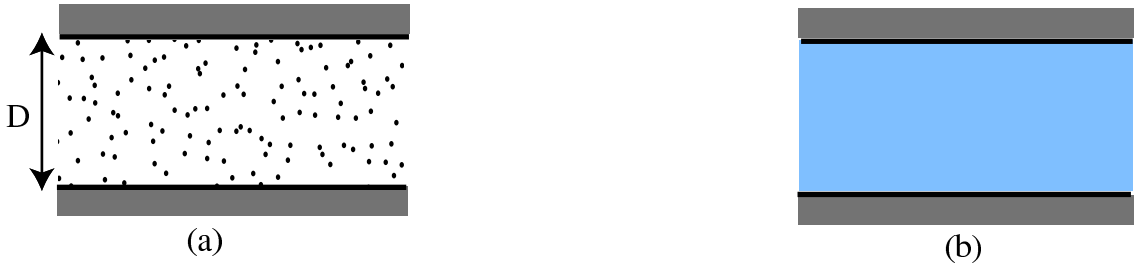}
\caption{$\Omega_{non-condensed}(\mu) = 2A \gamma_{sv}-DA P_v(\mu)
\qquad \Omega_{condensed} (\mu)= 2A \gamma_{sl}-DA P_l(\mu)$}
\label{fig:configurations}
\end{figure}

\bigskip

\bigskip

\noindent
with $\mu = \mu_{sat} - \Delta \mu$ the chemical potential of the reservoir.
Outside of coexistence, i.e.\ if $\Delta \mu \neq 0$, the pressure in the two phases are different  and
are given by the  thermodynamic relation $\partial (P_l-P_v)/\partial \mu = \rho_l-\rho_v$ (with $\rho_l$, $\rho_v$  the number of molecules per unit volume in each phase). As the liquid is usually much more dense and incompressible than the vapor the pressure difference reduces to
$(P_v-P_l )(\mu) \simeq \rho_l \Delta \mu = \rho_l k_BT {\rm ln} (P_{sat}/P_v)$
if the vapor can be considered as an ideal gas.
Thus, the condensed state is favored if the confinement is smaller than the critical distance $D_c(\mu)$:
\begin{equation}
\rho_l \Delta \mu D_c(\mu)= 2(\gamma_{sv}-\gamma_{sl})
\label{eq:criticaldistance}
\end{equation}
The left hand side of eq.\ (\ref{eq:criticaldistance}) represents the free energy required to condense the unfavorable liquid state, and the r.h.s.  the gain in surface energy.  $D_c(\mu)$ is thus the critical
distance which balances the surface interactions and the bulk interactions to determine the
phase diagram of the fluid (Israelachvili 1992).
From the above equation it is clear that capillary condensation can occur only if the liquid wets, at least partially, the solid surfaces.

\singlespacing

%%%%%%%% Insert A
\fbox{
\parbox{14.8cm}{
\begin{center}
\subsubsection*{Insert A: Surface tension and contact angle}
\end{center}
The surface tension of a fluid interface  is defined in terms of the work required to increase its area

$$\gamma_{lv}=\left ( \frac{\partial \mathcal F}{\partial A_{lv}} \right )_{N_l,V_l,N_v,V_v,T}$$
Here $\mathcal F$ is the free energy of a liquid-vapor system, $T$ its temperature, $A_{lv}$ the interface area, and $N_l$, $V_l$, $N_v$, $V_v$  the number of molecules  and the volume of each phase  respectively (Rowlinson and Widom 1982).
For a solid surface, one can likewise define a difference
of surface tension $\gamma_{sl}-\gamma_{sv}$ for the wet and dry surface in terms of the work
$d\mathcal F$ required to wet a fraction $dA_{sl}$ of the surface initially in the dry state.

It is shown in thermodynamics that the surface tension is  a grand canonical excess potential per unit area.
The total grand canonical potential of a multiphase system:

$$\Omega=-P_vV_v-P_lV_l-P_sV_s+\gamma_{lv}A_{lv}+ (\gamma_{sl}-\gamma_{sv}) A_{sl}$$
is the potential energy for an open system. Its variation is equal to the work done on the system during a transformation, and its value is minimal at equilibrium.

\smallskip

On the diagram of insert B let us consider a horizontal translation $dx$ of the meniscus. At equilibrium,
the grand canonical potential is minimum:
$ d\Omega  = -P_ldV_l-P_vdV_v+(\gamma_{sl}-\gamma_{sv})dA_{sl}= 0$.
Thus $P_v-P_l=(\gamma_{sv}-\gamma_{sl})/D$.
But, according to Laplace's law of capillarity, the pressure difference $P_v-P_l$ is also related to the curvature  of the meniscus:
$P_v-P_l=\gamma_{lv}/r=2 \gamma_{lv}{\rm cos}\ \theta/D$, where $\theta$ is the contact angle.
We deduce the Young-Dupr\'e law of partial  wetting:

\begin{equation}
\gamma_{lv}{\rm cos}\ \theta=\gamma_{sv}-\gamma_{sl} \qquad {\rm valid \ if }
\quad S= \gamma_{sv}- \gamma_{sl}- \gamma_{lv} \le 0
\label{eq:youngdupre}
\end{equation}
The parameter $S$ is  the wetting parameter (de Gennes \etal 2003). The situation $S>0$ corresponds to perfect wetting. In this case a thin liquid layer covers the solid surface (see insert C).
}}

%\doublespacing

\bigskip

\bigskip

In the case of partial wetting, the difference between the dry and wet surface tension is related to the contact angle $\theta$ of the liquid onto the solid surface and the critical distance reduces to

\begin{equation}
D_c(\mu)= \frac{ 2\gamma_{lv}{\rm cos} \theta}{\rho_l \Delta \mu}=2 r_K {\rm cos}\theta
\label{eq:distancecritique}
\end{equation}
where $r_K$ is the Kelvin's radius associated to the undersaturation $\Delta \mu$ (see insert B).

For an estimation of the order of magnitude of the confinement at which capillary condensation
occurs, consider the case of water at room temperature:  $\gamma_{lv}= 72$ mJ/m$^2$,
$\rho_l=5.5 \times 10^4$ mol/m$^3$, and assume a contact angle $\theta = 30^o$. In ambient conditions with a relative humidity of $P_v/P_{sat} = 40\%$, one has $r_k \simeq0.6$ nm and $D_c \simeq 1$ nm. The scale is in the nanometer range, and increases quickly with humidity: it reaches 4 nm at 80\% and 18 nm at 95\% relative humidity. Therefore, capillary condensates are ubiquitous in ambient condition in high confinement situations.

\bigskip

\singlespacing

%%%%%% Insert B %%%%%%%%%%
\fbox{
\parbox{14.8cm}{
\begin{center}
\subsubsection*{Insert B: Laplace-Kelvin equation}
\end{center}
Another way to address capillary condensation is to consider the coexistence of a liquid and its vapor  across a curved interface.
Because of the Laplace law of capillarity the pressure in the two phases are not equal:
$P_{int}-P_{ext}=\gamma_{lv}/r$,
with $r$ is the  radius of mean curvature of the interface. The pressure is always
higher on the concave side.
Because of this pressure difference the chemical potential of coexistence is shifted :

$$\mu_v(P_v)=\mu_l(P_l=P_v-\gamma_{lv}/r) = \mu_{sat}-\Delta \mu$$
We have assumed here that the liquid is on the convex side, a configuration compatible with an under-saturation. For an ideal vapor and an uncompressible liquid:
\begin{equation}
\Delta \mu = k_BT {\rm ln}(P_{sat}/P_v) \qquad P_l(\mu) \simeq \rho_l \Delta \mu+P_v(\mu)
\label{eq:gazparfait}
\end{equation}
from where we get the Laplace-Kelvin equation for the equilibrium  curvature (Thomson, 1871):
\begin{equation}
\frac{\gamma_{lv}}{r_K}=P_v(\mu)-P_l(\mu)\simeq \rho_l \Delta \mu= \rho_l k_BT {\rm ln}\frac{P_{sat}}{P_v}
% r_K (\Delta \mu)= \frac {\gamma_{lv}}{P_v(\mu)-P_l(\mu)}\simeq \frac{\gamma_{lv}}{\rho_l \Delta \mu}
\label{eq:LaplaceKelvin}
\end{equation}
\begin{center}
\parbox{8cm}{We check that in a flat slit, the critical confinement and the Kelvin's radius are related by
$D_c=2 r_K {\rm cos} \theta$. The capillary condensate is thus limited by a meniscus whose curvature is equal to the Kelvin's radius. The Laplace-Kelvin law is however more general than eq.\ (\ref{eq:distancecritique}) and allows to predict the critical confinement in arbitrarily complex geometries.}
\parbox{5mm}{.}
\parbox{6cm}{\includegraphics{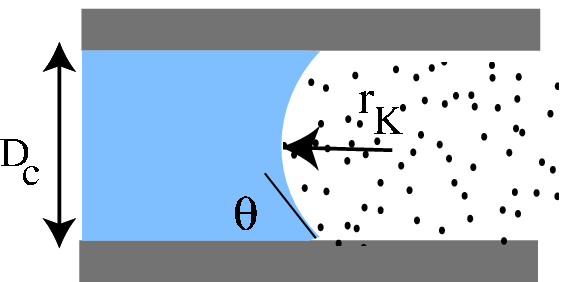}}
\end{center} }}
%%%%%%%% FIN Insert B%%%%%%%%%

%\doublespacing

\bigskip

\bigskip

We see from the Laplace-Kelvin equation that the pressure in capillary condensates is usually very low : taking the example of water in ambient conditions with relative humidity $P_v/P_{sat}=40\%$, the pressure in the condensates is $P_l=-120$ MPa, i.e.\ -1200 bar. With these severe negative pressures, condensates exert strong attractive capillary forces on the surfaces to which they are adsorbed. Thus capillary condensation is usually associated to important mechanical aspects, such as cohesion, friction, elastic instabilities and micro-structures destruction.
Furthermore, if the liquid phase wets totally the solid surfaces  (see insert A), the
surfaces may be covered by a liquid  film even in a non-confined geometry  (see inserts C and D). In this case the critical distance for capillary condensation can be significantly enhanced at low humidity. In the case of water, the condensation of a liquid film has important consequences on surface chemistry as surface species can be dissolved  in the liquid phase, and the capillary condensation at the level of contact between surfaces increases solute transport and is responsible for dissolution-recrystallisation processes which lead to slow temporal evolution of materials mechanical properties (cf.\ sect.\ \ref{sec:SurfaceChemistry}).

%\bigskip

\subsection{Mesoporous systems}

\begin{figure}
\centering
\includegraphics[width=8cm]{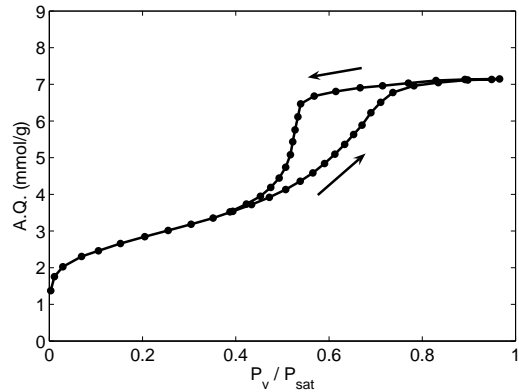}
\caption{Sorption isotherm of nitrogen at 77 K in Vycor (data from Torralvo \etal 1998, A.Q. = adsorbed quantity).}
\label{fig:isotherme}
\end{figure}

Capillary condensation has been extensively studied in relation to sorption isotherms in mesoporous media - i.e. nanomaterials with pore sizes between 2 and 50 nm - in the prospect of using  those isotherms   for the determination of porosity characteristics such as the specific area and the pore size distribution.
Fig.\ \ref{fig:isotherme} for instance shows a typical adsorption isotherm of nitrogen in a mesoporous silica at 77 K. In a first domain of low vapor pressure, the adsorption is a function of the relative vapor saturation only, and corresponds to the mono- and polylayer accumulation of nitrogen on the solid walls. This regime allows the determination of the specific area, for instance  through the Brunauer-Emmett-Teller model (1938). At a higher pressure  a massive adsorption corresponds to capillary condensation, and the porous volume is completely filled by liquid nitrogen before the saturating pressure is reached. This adsorption branch shows usually a strong hysteresis and the capillary desorption is obtained at a lower vapor pressure  than the condensation. This feature underlines the first order nature of capillary condensation. It is shown in the next paragraph that for sufficiently simple pore shapes the desorption branch is the stable one and corresponds to the liquid-vapor equilibrium through curved menisci.  The desorption branch may be used to determine the pore size distribution of the medium through the Laplace-Kelvin relation using appropriate models (Barret-Joyner-Halenda 1951). More can be found on the physics of phase separation in confined media in the review of Gelb \etal (1999).

\smallskip

\singlespacing

%%%%%%    insert C %%%%%%%%
\fbox{
\parbox{14.8cm}{
\begin{center}
\subsubsection*{Insert C: Perfect wetting: the disjoining pressure}
\end{center}
%\centerline{Perfect wetting : the disjoining pressure}
When the energy of the dry solid surface $\gamma_{sv}$ is larger than the
sum $\gamma_{sl}+\gamma_{lv}$   of the solid-liquid and liquid-vapor interfaces ($S >0$),
the affinity of the solid for the fluid is such that it can stabilize a liquid film
in equilibrium with an undersaturated vapor  without any  confinement.
The existence of such wetting  films must be taken
into account when determining the liquid-vapor equilibrium in a confined space.

\bigskip

\begin{center}
\includegraphics[width=6cm]{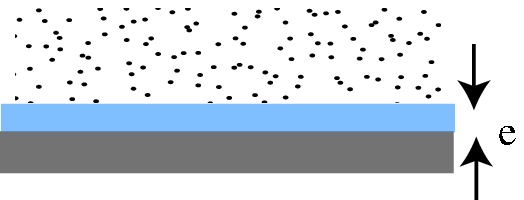}
\label{fig:films}
\end{center}
In the theory of wetting, liquid films are described by the concept of interface
potential (Derjaguin 1955, de Gennes 1985). The excess potential per unit area of a solid surface covered by a wetting film
does not reduce to the sum $\gamma_{sl}+\gamma_{lv}$ of the surface tensions: a further excess
must be taken into account corresponding to the fact that the molecular
interactions which generate the surface tension do not operate over a thickness
of liquid  that can be considered infinite. The excess grand
canonical potential of the humid solid surface of area $A$ is then
\begin{equation}
\frac{\Omega_{sv}}{A} =\tilde \gamma_{sv}= \gamma_{sl}+\gamma_{lv} + W_{slv}(e)-e(P_l(\mu)-P_v(\mu))
\label{eq:omegainterf}
\end{equation}
where the interface potential $W_{slv}(e)$ vanishes for a macroscopic film.
%and takes the value $W_{slv}(0)=   \gamma_{sv}- \gamma_{sl}-\gamma_{lv}$ for a dry solid.
The  excess potential $\Omega_{sv}/A$ describes the thermodynamic
properties of the liquid film. It is minimum at equilibrium,  so that the pressure in the liquid  is not the
same as in the vapor:
\begin{equation}
P_v(\mu)-P_l(\mu)=  -\frac{dW_{slv}(e)}{de}=\Pi_d
\label{eq:disjoiningpressure}
\end{equation}
The pressure difference $\Pi_d$  is called the disjoining pressure. The interface potential $W_{slv}(e)$ and the wetting parameter $\tilde \gamma_{sv}(\Pi_d)-\gamma_{sl}-\gamma_{lv}$ are Legendre transforms of each other:
\begin{equation}
%\tilde \gamma_{sv}(\Pi_d)=\gamma_{sl}+\gamma_{lv} + W_{slv}(e)+e\Pi_d \quad
W_{slv}(e) =\tilde \gamma_{sv}(\Pi_d)-\gamma_{sl}-\gamma_{lv}-e\Pi_d
%\frac{\partial \tilde \gamma_{sv}}{\partial \Pi_d}
\quad e=\frac{\partial \tilde \gamma_{sv}}{\partial \Pi_d}
\label{eq:legendre}
\end{equation}
For instance in the case of van der Waals forces, the interface potential results from dipolar
interactions going as $1/r^6$ between molecules, and varies as $1/e^2$:
\begin{equation}
W_{slv}(e)=-\frac{A_{slv}}{12\pi e^2} \quad
\Pi_d(e)= -\frac{A_{slv}}{6\pi e^3} \quad
\tilde \gamma_{sv} = \gamma_{sl}+\gamma_{lv} + \left ( \frac{-9A_{slv}}{16 \pi}\right ) ^{1/3} \Pi_d^{2/3}
\label{eq:hamaker}
\end{equation}
The Hamaker constant $A_{slv}$ has the dimension of an energy (Israelachvili 1992). It lies typically between $10^{-21}$ and $10^{-18}$ J and  has negative sign when the liquid wets the solid, i.e. if the interface potential is positive.}}
%%%%%%%% FIN Insert C%%%%%%%%%

%\doublespacing
%\singlespacing

%%%%%%%%% Insert D%%%%%%%%%%%
\fbox{
\parbox{14.8cm}{
\begin{center}
\subsubsection*{Insert D: Perfect wetting: the prewetting transition and capillary condensation}
\end{center}
In a situation of perfect wetting,  a liquid film condenses on a flat  isolated  solid surface if
the humid solid surface tension $\tilde \gamma_{sv} $  is lower than the dry one:
\begin{equation}
\tilde \gamma_{sv} =\gamma_{sl}+\gamma_{lv} + W_{slv}(e) +e \Pi_d \le \gamma_{sv}
\label{eq:prewetting}
\end{equation}
If the film exists, its  thickness  at equilibrium with the vapor   is implicitly determined by the analogue of the Laplace-Kelvin equation (\ref{eq:LaplaceKelvin}):
\begin{equation}
\Pi_d(e)=-\frac{\partial W_{slv}(e)}{\partial e}=P_v(\mu)-P_l(\mu) = \rho_lk_BT \ {\rm ln} \frac{P_{sat}}{P_v}
\label{eq:DisjoiningKelvin}
\end{equation}
The thickness $e^*$ realizing the equality in relation (\ref{eq:prewetting}) is a minimum thickness for the wetting film, and the associated chemical potential $\mu^*$ and vapor pressure $P_v^*$ correspond to a {\em prewetting transition}.
Above the transition  the  thickness of the adsorbed film  increases with the vapor pressure until it reaches a macroscopic value  at saturation. In the case of van der Waals wetting for instance the vapor pressure at the prewetting transition is given by
$\Pi_d^* = \rho_lk_BT {\rm ln} (P_{sat}/P_v^*) = \sqrt{16 \pi S^3/(-9A_{slv})}$
%$$\rho_lk_BT {\rm ln} \frac{P_{sat}}{P_v^*} = \frac{4}{3} \sqrt{\frac{\pi S^3}{-A_{slv}}}
%\qquad {\rm with}\quad S=\gamma_{sv}-\gamma_{sl}-\gamma_{lv}$$
with $S$ the wetting parameter (\ref{eq:youngdupre}).

\medskip

In a confined geometry such as sketched in Fig.\ \ref{fig:configurations}, the grand canonical
potential of the ``non-condensed'' state is shifted above the prewetting transition as the solid surface tension $\gamma_{sv}$ has to be replaced by the humid value $\tilde \gamma_{sv}$.
The modified eq.\ (\ref{eq:criticaldistance})
and the Laplace-Kelvin relation $\gamma_{lv}/r_K=\Pi_d=\rho_l\Delta \mu$ give the critical distance
(Derjaguin and Churaev 1976):
\begin{equation}
D_c =2 r_K+2e+2\frac{W_{slv}(e)}{\Pi_d}
\label{eq:Dcmouillant}
\end{equation}
The difference with the partial wetting case  is not simply to decrease the available interstice by twice the film thickness.
In the case of van der Waals forces for example,
\begin{equation}
D_c=2r_K+3e \qquad {\rm with} \quad e=(-A_{slv}/6\pi  \rho_l \Delta \mu)^{1/3}
\label{eq:3e}
\end{equation}
The effect of adsorbed films become
quantitatively important for determining the critical size for capillary condensation in a perfect wetting situation.
}}
%%%%%%%% FIN Insert D%%%%%%%%%

%\doublespacing

\newpage

\section{Capillary adhesion Forces}
\label{sec:CapillaryForces}

\subsection{Measurements by SFA}
\label{sec:MeasureForceSFA}

Because of their high curvature, capillary condensates exert a large attractive force on the surfaces they connect. In return these large forces represent a valuable tool  to study their thermodynamic and mechanical properties. Experimentally, the ideal geometry involves a contact with at least one curved surface - either a sphere on a plane, two spheres or two crossed cylinders - so that locally the topology resumes to  a sphere of radius $R$ close to a flat. Surface Force Apparatus (SFA) use macroscopic radius $R$ in order to take advantage of the powerful Derjaguin approximation, which relates the interaction force $F(D)$ at distance $D$ to the free energy per unit area (or other appropriate thermodynamic potential)  of two flat parallel surfaces at the same distance $D$ (see insert E). It must be emphasized that  the Derjaguin approximation accounts exactly for the contribution of the Laplace pressure, and more generally for all ``surface terms'' contributing to the force, but it does not account properly for the ``perimeter terms'' such as the line forces acting on the border of the meniscus, so that it neglects terms of order $\sqrt{r_K/R}$.
\medskip

\singlespacing

\fbox{
\parbox{15.2cm}{
\begin{center}
\subsubsection*{Insert E: Derjaguin approximation}
\end{center}
%\begin{center}
 \parbox{8cm}{Consider a sphere of macroscopic radius $R$ at a distance $D<<R$ from a plane. If the range of the  surface interactions   is also smaller than $R$, then all surface interactions  take place in the region  were the sphere is almost parallel to   the plane and the global interaction force $F(D)$ is:}
  \parbox{5mm}{\ }
 \parbox{7.2cm}{\includegraphics{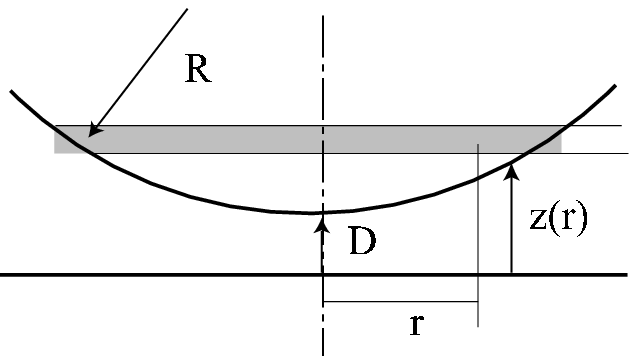}}
% \end{center}
$$F(D) = -\frac{d}{dD} \left ( \int _D^\infty dA(z) [U(z)-U(\infty)] \right )
 = -2\pi R \frac{d}{dD} \int _D^\infty  [U(z)-U(\infty)] dz $$
with $dA(z)\simeq 2\pi R dz$ the sphere area  between altitude $z$ and $z+dz$, and $U(z)$ the appropriate potential energy per unit area for flat, parallel surfaces at distance $z$.
$$ {\rm This\  gives\  the\  Derjaguin\  approximation:} \quad
F(D)=2\pi R \left [U(D)-U(\infty) \right ]$$
}}
\medskip

\bigskip

%\doublespacing

In a surrounding condensable vapor the appropriate potential is the grand potential per unit area considered in Fig.\ \ref{fig:configurations}:
\begin{equation}
\Omega (D<D_c) = (\rho_L-\rho_v) \Delta \mu D + 2\gamma_{sl} + W_{sls}(D)
\qquad \Omega (D>D_c) = 2\gamma_{sv} +W_{svs}(D)
\label{eq:omegapourforce}
\end{equation}
with $W_{sls}(D)$ (resp. $W_{svs}(D)$)  the solid-solid interaction through the liquid phase (resp. vapor phase). This  gives the interaction force in the case of partial wetting:
\begin{equation}
F(D<D_c)=2\pi R \left [ -2 \gamma_{lv} {\rm cos} \theta + \frac{\gamma_{lv}}{r_K} D  \right ]+ 2\pi RW_{sls}(D)
\label{eq:capforce}
\end{equation}
and the pull-off force (Fisher \etal 1981) :
\begin{equation}
F_{pull-off}= - F(D=0) = 4 \pi R \gamma_{lv}{\rm cos} \theta +  4 \pi R\gamma_{sl} = 4\pi R \gamma_{sv}
\label{eq:pull-off-partial}
\end{equation}
It should be mentioned here that the prefactor $4\pi R$ entering eq.\ (\ref{eq:pull-off-partial}) is valid only for rigid surfaces, i.e. if the Tabor parameter $\lambda = ( \gamma_{lv} R^2/ E^{*2}h_o^3)^{1/3}<0.1$ with $E^{*}$ the (mean) reduced Young modulus of the solid, and $h_o$ the range of the interaction.  In the general case  the $4\pi R$ prefactor in eq.\ (\ref{eq:pull-off-partial}) should be replaced by $\epsilon(\lambda) \pi R$   with $3 \le \epsilon \le 4 $ (Maugis 1992, Maugis \etal 1993).

\begin{figure}
 \centering
 \parbox{7cm}{\includegraphics[width=7cm]{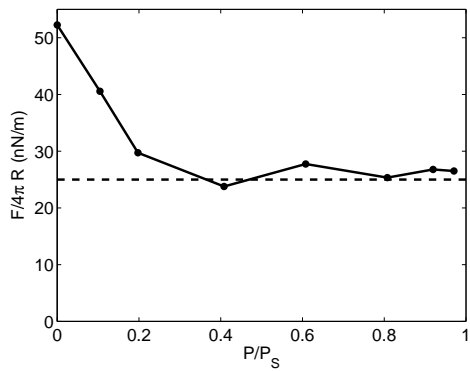}}
 \caption{Mesured pull-off force as a function of the relative vapor pressure of cyclohexane. The dashed line is the bulk surface tension of cyclohexane ($\gamma_{lv} = 25$ mN/m; data after Christenson 1988).}
\label{fig:SFAChrist}
\end{figure}

The pull-off force (\ref{eq:pull-off-partial}) is the sum of the capillary force $4 \pi R \gamma_{lv}{\rm cos} \theta$ and of the solid-solid adhesion in the liquid phase $4 \pi R \gamma_{sl}$.  A remarkable feature of eq.\ (\ref{eq:pull-off-partial}), valid in a {\em partial wetting} situation as we should recall, is that  the capillary force does not depend on the volume of the liquid bridge nor on its curvature, a property due to the fact that the wetted area and the capillary pressure exactly compensate each other when the size of the condensate varies. Furthermore  the pull-off force does not even depend on the presence of the condensate and is identically equal to its value in dry atmosphere, $4\pi R \gamma_{sv}$.  Of course these remarkable results require a perfectly smooth  geometry. In practice, even minute surface roughness screens very efficiently the  solid-solid interactions, so that it can be neglected only in condensable vapor close to saturation. A major experimental problem is then to follow the transition of the pull-off force from $4\pi R\gamma_{sv}$ to $4 \pi R (\gamma_{lv}{\rm cos} \theta +  \gamma_{sl})$ as the relative vapor pressure varies from 0 to 1.
Surface Force Apparatus have addressed this problem using atomically smooth mica surfaces glued on crossed cylinders.  However even with these physically ideal surfaces the pull-off force measured with dry surfaces gives a  solid-solid surface tension $\gamma_{sv}$ significantly lower than the one obtained, for instance,  from cleavage experiments
or contact angle measurements (Christenson 1988). The same occurs in liquid phase, where a monolayer of liquid remains stuck between the mica surfaces and prevents the measurement of the actual solid-liquid surface tension. Therefore a full  experimental verification of eq.\ (\ref{eq:pull-off-partial}) over the whole vapor pressure range is not available. It is found however, that the pull-off force is dominated by the capillary force and that it does not depend on the vapor pressure over a significant range including saturation (cf.\ Fig.\ \ref{fig:SFAChrist}) (Fisher \etal 1981, Christenson 1988).

As the capillary force is less sensitive to roughness than the direct solid-solid adhesion (cf.\ insert H), SFA pull-off forces coupled to direct measurements of the Kelvin's radius via the observation of the liquid bridge extension have been used to investigate the validity of the macroscopic capillarity at nano-scale, and an eventual drift of the liquid surface tension for highly curved menisci  (Fisher \etal 1981).
The remarkable results is that  the Laplace law of capillarity  holds down to very small radii of curvature (i.e. less than 2-3 nm) for simple liquids. Discrepancies have been reported with water, but they have been shown to depend on the type of  ions  covering the mica surfaces, and have been attributed to the accumulation of involatile material in the water condensate (Christenson 1985, Christenson 1988).

\smallskip

Valuable informations have been obtained by studying the full profile of the capillary force  instead of only the pull-off force, as the slope of  $F(D)$ gives a direct measure of the Laplace pressure $\gamma_{lv}/r_K$. This requires however a SFA of large stiffness.  Such analysis has been performed by Crassous \etal (1993) in a {\em perfect wetting} situation.
In this case  the theoretical  force is obtained from eq.\ (\ref{eq:omegapourforce}) using the reference $\Omega(D=\infty)=2 \tilde \gamma_{sv}$ instead of $2 \gamma_{sv}$ (Charlaix \etal 2005):
\begin{equation}
F(D<D_c)=2\pi R \left [\frac{\gamma_{lv}}{r_K} D  +2 \gamma_{sl} -2 \tilde \gamma_{sv} \right ] +2\pi R W_{sls}(D)
\label{eq:wettingforce}
\end{equation}
The first term on the r.h.s. is the capillary force at liquid-vapor equilibrium. It  also writes $F_{cap}=2\pi R \Pi_d(D-D_c)$ with $D_c$ given by eq.\ (\ref{eq:Dcmouillant}).  Its slope   provides the Laplace/disjoining pressure $\gamma_{lv}/r_K=\Pi_d$, while the offset gives access to $\tilde \gamma_{sv}$ and to the wetting potential with the help of eq. (\ref{eq:legendre}). Crassous \etal verified experimentally eq.\ (\ref{eq:wettingforce}) down to Kelvin's radii of 3 nm  in the case of heptane condensation on platinum surfaces.

\begin{figure}
 \parbox{10cm}{\includegraphics[width=10cm]{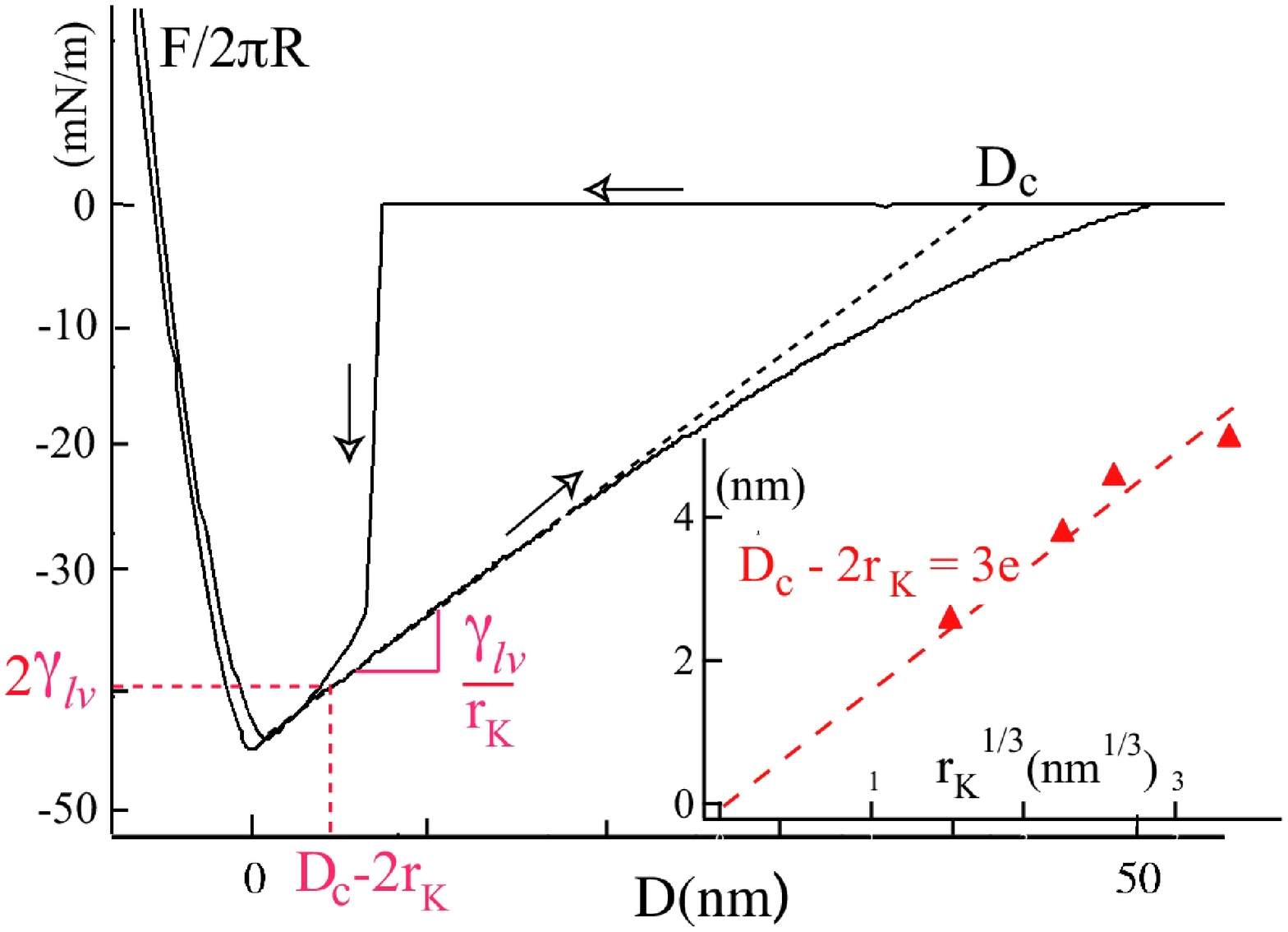}}
 \parbox{7cm}{\includegraphics[width=7cm]{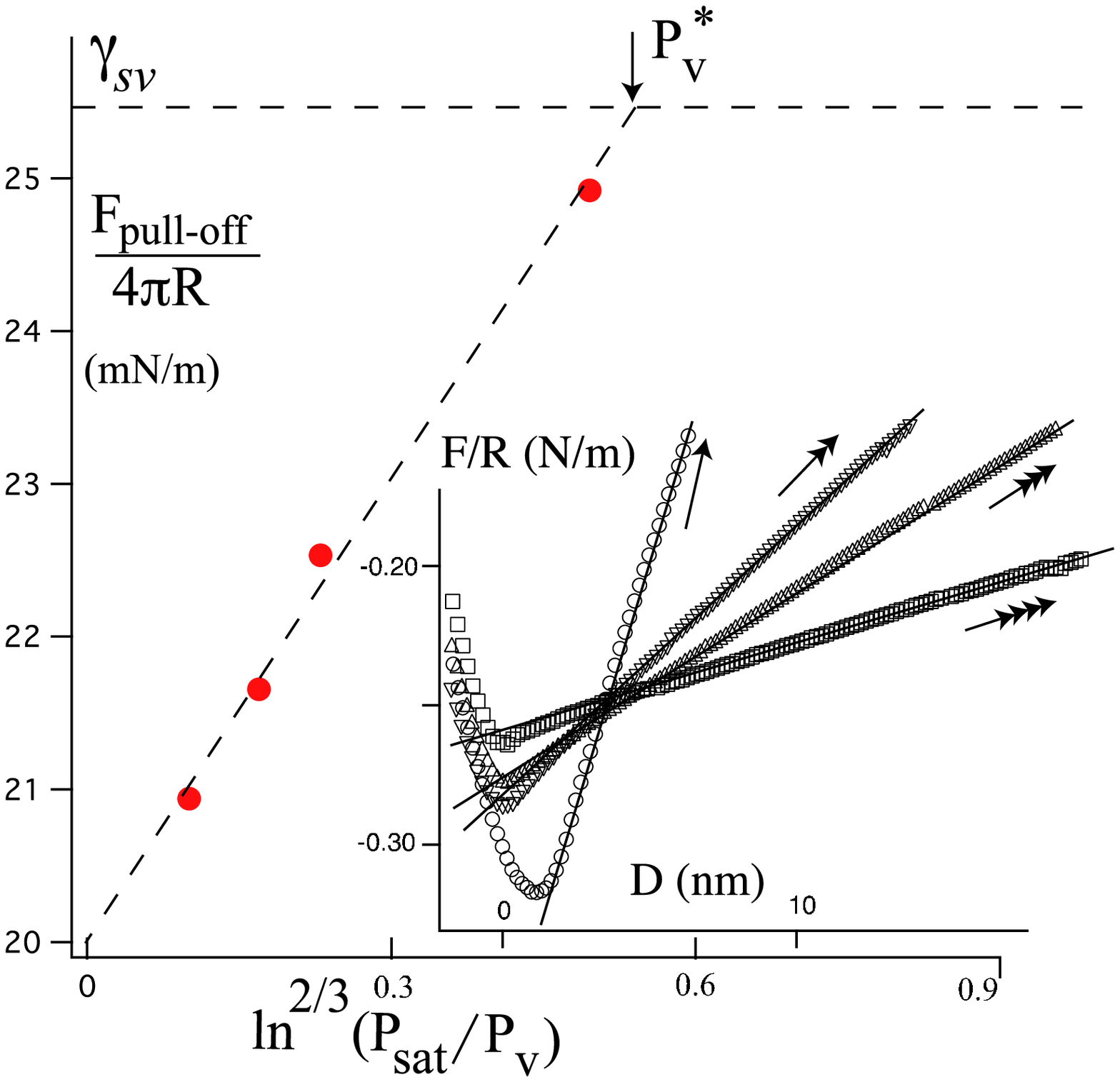}}
 \caption{Left: the capillary force between platinum surfaces measured in n-heptane vapor. The attractive force varies linearly with the distance on a significant range where the condensate is at equilibrium with the vapor. For van der Waals forces the thickness $e$ of the wetting film covering the surfaces is obtained through $F_{cap}(3e)=4\pi R \gamma_{lv}$. It varies as the power $1/3$ of the Kelvin's radius. Right: the pull-off force measured for various vapor pressures, and the comparison with the van der Waals dependence (\ref{eq:VanderWaalsPulloff}). The pull-off force is expected to saturate to the dry pull-off force at the prewetting transition.}
\label{fig:SFA}
\end{figure}

A consequence of eq.\ (\ref{eq:wettingforce}) is to give the  pull-off force as a function of the vapor saturation
\begin{equation}
F_{pull-off}=4\pi R \tilde \gamma_{sv}
\label{eq:pull-off-total}
\end{equation}
with $\tilde \gamma_{sv}(\Pi_d)$ given by (\ref{eq:legendre}) and $\Pi_d$ by (\ref{eq:DisjoiningKelvin}).
For instance with van der Waals forces one expects:
\begin{equation}
F_{pull-off}=4\pi R \left [  \gamma_{sl}+\gamma_{lv} + a \left (  {\rm ln}
\frac{P_{sat}}{P_v} \right ) ^{2/3} \right ]
\label{eq:VanderWaalsPulloff}
\end{equation}
with $a=(-9A_{slv}\rho_l^2 (k_BT)^2/16 \pi)^{1/3}$. When the vapor pressure is increased from $0$ to $P_v^*$, the pull-off force remains equal to the dry value $4\pi R \gamma_{sv}$. Above $P_v^*$ the pull-off force {\em decreases} with increasing vapor pressure, until it reaches the wet limit $4\pi R (\gamma_{sl}+\gamma_{lv})$.

\smallskip

A major feature of capillary condensation is the strong hysteresis in the formation  of the condensates.
As shown on Fig.\ \ref{fig:SFA} for instance, the non-condensed state is highly metastable for $D<D_c$, and an energy barrier has to be overcome to initiate the condensation process. The kinetics of capillary condensation has been much less studied than the static properties of the condensate.
Several theoretical works have focused on the calculation of the energy barrier which depends on the confined geometry (Restagno \etal 2000, Lefevre \etal 2004). In the case of a wetting situation and of an extended slit like in SFA experiments, it has been  shown  that the distance of condensation $D_s$ (see Fig.\ \ref{fig:SFA}) corresponds to the spinodal instability of the wetting film covering the solid surfaces (Derjaguin and Churaev 1976, Christenson 1994, Crassous \etal 1994). The growth kinetics of  the capillary bridge after its initial formation has been shown by Kohonen \etal (1999) to be reasonably well described by a model based on the diffusion-limited flow of vapor towards the bridge,
except for water condensates  which grow significantly more slowly; which is attributed to the effect of dissolution of inorganic material from the mica surfaces.

\bigskip

\subsection{Measurements by AFM}

The great relevance of capillary forces between nanoscale objects has stimulated the application of contact AFM to provide a more quantitative measurement of the forces acting between the very sharp probe tips (with typical radius of curvature of few nm) and samples with variable compositions or textures. The great advantage of AFM %are the extreme sensitivity to small forces and
is the elevated lateral resolution, which is comparable with the curvature radius of the terminal portion of the probe tip. The first order effect of the probe size reduction is a proportional reduction of the adhesion forces to typical values in the nN range, which can easily be measured by AFM pull-off tests, i.e.\ by measuring the force of contact break down during a force-distance (approach-retract) curve. However, the typical cantilever stiffness (in the range of a few N/m) prevents the full characterization of the $F(D)$ curve that can be obtained by SFA (cf.\ previous section), and typical data are limited to the variation of the pull-off force with relative humidity, corresponding to $D=0$ (Xiao and Qian 2000).

Before entering a deeper analysis, we should remark that the AFM force measurements are very sensitive to many ill controlled factors, the more important being the exact details of the tip shape, including its nanoscale roughness. Although continuum derived laws become questionable at scales smaller than 1-2 nm, we will first establish how the capillary force at a finite distance $D$ should properly be evaluated for nanoscale probes according to classical modeling and then discuss the limitations of this approach.

\begin{figure}[!h]
\centering
\includegraphics[width=0.45 \textwidth]{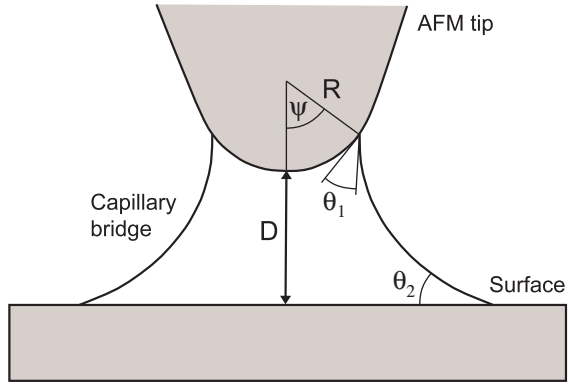}
\caption{Graphical representation of a capillary bridge formed between an AFM tip of radius $R$ and a flat surface separated by a distance $D$. $\psi$ is the filling angle, $r_1$ and $r_2$ are the principal radii of curvature of the meniscus, $\theta_1$ and $\theta_2$ are the contact angles of the AFM tip and surface.}
\label{fig:CapBridgeAFM}
\end{figure}

Let the AFM tip be modeled as a cone terminating with a spherical tip of radius $R$ and contact angle $\theta_1$ and the sample as a flat surface of contact angle $\theta_2$ (cf.\ Fig.\ \ref{fig:CapBridgeAFM}). The Derjaguin approximation (insert E) leading to eqs.\ (\ref{eq:pull-off-partial}) and (\ref{eq:pull-off-total}) is no longer applicable since the $r_K/R$ ratio is not negligible for nanoscale probes.
Although other approximations exist, such as the circular meniscus surface model, their domain of validity is limited and their inattentive use can lead to severe errors. We thus recommend to refer to the complete analytical solutions provided by Orr \etal (1975), that describe constant mean curvature meniscus surfaces for systems amenable to a generic sphere/plane geometry (insert F), along with a computation of capillary forces.
A special attention must be paid to the case of perfect wetting, since the Orr equations do not take into account the presence of wetting films. This situation was properly treated by Crassous (1995) using both an approximated solution and numerical simulation. In the case of a vanishing distance $D$, the solid-solid interactions should also be accounted for.

The capillary force $F_C$ acting on the sphere has two main contributions: a capillary pressure force $F_P$, which is related to the negative Laplace pressure in the liquid bridge and acts on the sphere surface, and a surface tension force $F_S$, which resides in the meniscus and acts on the contact line between the sphere and the liquid bridge. While this second term is negligible in the Derjaguin approximation, when $R$ decreases to nanoscale its contribution becomes of the same order as the Laplace pressure term, since the ratio of the perimeter to surface of the capillary meniscus contact is increased. These terms can be written without any approximation as:
\begin{equation}
F_C = F_P + F_S \qquad F_P = -2\pi \gamma H R^2 \sin^2\psi \qquad F_S = 2 \pi \gamma R \sin \psi \sin(\theta_1 + \psi)
\label{eq:FCapForceTerms}
\end{equation}

The filling angle $\psi$ should then be related to relative humidity by using the Orr \etal relations to evaluate the surface mean curvature (insert F) and by substituting it into the Laplace-Kelvin relation (\ref{eq:LaplaceKelvin}).

\bigskip

\singlespacing

%%%%%%%% FICHE 6
        \fbox{
\parbox{14.8cm}{
\begin{center}
\subsubsection*{Insert F: Orr solutions for negative curvature menisci}
\end{center}

As observed by Plateau (1964), the meridian meniscus profile of a capillary bridge is expected to step through a series of shapes, including portions of nodoids, catenoids, unduloids, cylinders and spheres, as a function of the systems geometrical and wetting parameters.
A clear report on the meniscus shapes as a function of the constant mean curvature can be found in Orr \etal (1964), generally expressed in terms of elliptic integrals, as well as a characterization of the capillary forces that are the focus here. However, only nodoids are appropriate for describing capillary bridges equilibrated with undersaturated humidity, and catenoids are limited to the case of vanishing curvature (corresponding to saturated humidity or contact with a bulk liquid reservoir). We will thus only discuss here the case of a negative curvature. The mean curvature $H = 1/2r_K<0$ can be related to the parameters $R$, $D$, $\theta_1$, $\theta_2$ and $\psi$ by:
\begin{equation}
2HR/\Psi = \Theta - {1 \over k} [E(\phi_2,k)-E(\phi_1,k)] + {1 - k^2 \over k} [F(\phi_2,k) - F(\phi_1,k)]
\label{eq:OrrCurvature}
\end{equation}
where
$$\Psi = {1 \over D + 1 - \cos \psi} \qquad \Theta = -\cos(\theta_1 + \psi) - \cos \theta_2$$
$$\phi_1 = -(\theta_1+\psi) + {\pi \over 2} \qquad \phi_2 = \theta_2 - {\pi \over 2} \qquad k = {1 \over (1+c)^{1/2}}$$
$$c = 4 H^2 R^2 \sin^2 \psi - 4 H R \sin \phi \sin(\theta_1 + \psi)$$
and the functions $F$ and $E$ are the incomplete elliptic integrals of first and second type (Abramowitz and Stegun, 1972).
Since the curvature $H$ is present on both left and right hand terms of eq.\ (\ref{eq:OrrCurvature}), an iterative procedure is needed for numerical solution. The mean curvature $H$ should then be related to the relative humidity through the Laplace-Kelvin relation (\ref{eq:LaplaceKelvin}). The contribution of capillary forces to the pull-off force can then be simply obtained by the taking the maximum value of the total force, which is obtained for $D=0$.
}}
%%%%%%%% FIN Insert F
%\doublespacing

\bigskip

\begin{figure}[!h]
\centering
\includegraphics[width=7 cm]{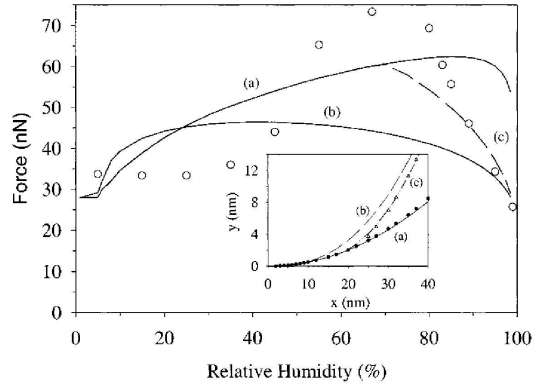}
\caption{Adhesion force (pull-off) measurements between a $Si_3N_4$ AFM tip and a $SiO_2$ surface (circles). The theoretic approximated expressions calculated from the equations of de Lazzer \etal (1999) are also shown for the three different tip shapes with radial profile $y(x)$ represented in the insert (overlapping for $x<10$ nm): (a) parabolic profile $y = k_1 x^2$, (b) $y = k_2 x^{2.5}$, and (c) more complex dull shape (Figure after Xiao and Qian 2000).}
\label{fig:XiaoAdhesion}
\end{figure}

However, the applicability of this subtle modeling to AFM pull-off force measurements is presently hampered by several technical problems. On the opposite of SFA, the shape of typical AFM probes is generally ill-defined and subject to easy degradation and contamination.
Fig.\ \ref{fig:XiaoAdhesion} shows a typical example of pull-off force measurements as a function of humidity (Xiao and Qian 2000), along with the approximated modeling by de Lazzer \etal (1999) aimed at showing the effect of a more general tip shape in the form of a revolution solid with radial profile $y(x)$. Even if the considered tip profiles (cf.\ insert in Fig.\ \ref{fig:XiaoAdhesion}) are quite similar, the modeled curves are quite different and can not account for the measured data.
Although this modeling is based on the circular meniscus approximation and on an approximated estimate of the solid-solid van der Waals interactions, it can be retained as a clear proof of the critical effect of the tip shape on the capillary forces
(cf.\ also Pakarinen \etal 2005 for a less approximated modeling of the same data, leading to similar conclusions).
It must be acknowledged that articles on adhesion measurements from the AFM community often do not show the same maturity as the SFA community in the subtle modeling of these phenomena, and thus often contain misleading interpretations that delay the establishment of a strong based modeling.

The physical interpretation of AFM adhesion measurements in moist air still needs much work. The data in Fig.\ \ref{fig:XiaoAdhesion} can be considered as quite typical. The presence of a plateau at low humidity was often interpreted as the consequence of the non formation of a capillary bridge (Xiao and Qian 2000), reducing the adhesion force to the dry van der Waals solid-solid interaction. However, we showed in section \ref{sec:MeasureForceSFA} that no changes in the total adhesion force should be observed in the case of a vanishing meniscus between a perfect sphere and a plane in contact: even if the capillary bridge is strictly disappearing, the increased contribution of unscreened solid-solid interactions should strictly compensate the total adhesion force. The low humidity plateau could also be related to the effect of a small scale roughness in the tip and substrate (cf.\ insert H). At low humidity only the small contacting asperities would contribute to the pull-off force, and the rising regime is likely to be related to a transition towards a monomeniscus that may be treated according to the modeling presented in this section.
More insights into the behavior at low humidity will be provided in section \ref{sec:friction} by investigating the effect of capillary forces on cohesion and friction forces. The decreasing trend at higher humidity can be explained according to eq.\ (\ref{eq:FCapForceTerms}):
for small values of $R$ (less than 1 $\mu$m), the increase of the meniscus section with humidity is not compensating any more the decrease of the Laplace pressure in the pressure term $F_P$, due to the large values of the filling angle $\psi$ . We remark that although the surface tension term $F_S$ is significantly increasing at elevated humidity, the decrease of the pressure term has a dominant effect on the total capillary force.

A major subject of open debate is the limit of validity of the continuum modeling at nanoscale, for both the capillary forces and their combination with the other relevant interactions such as van der Waals, electrostatic interactions, hydrogen bonding and other kinds of specific chemical interactions between the contacting tip and substrate. Although the capillary forces are generally dominant in the adhesion force (due the efficient screening of other interactions), this is not the case at low humidity where the solid-solid interactions become more relevant, especially for small radius probes. A notable example is given by the strong impact of electrostatic interactions in the adhesion force in dry atmosphere (Wan \etal 1992).
Although the continuum derived laws were shown to describe the adhesion forces in SFA measurements down to scales as small as 1-2 nm (Lefevre \etal 2004), the discrete nature and structure of solid and liquid molecules should become relevant at subnanometer scales, where the phase behavior of water is still debated. Several investigators have reported the complexity of the structure of thin water films, especially at low humidity where ice-like behavior is expected (cf.\ Verdaguer \etal 2006). Moreover, Xu \etal (1998) have shown that the contact with an AFM tip can subtly modify the structure of the water layer on a mica substrate: 2D islands are left on the surface after removal of the tip and can persist several hours before evaporating. However, a significant disagreement is still found in literature on the details of the humidity dependence of these behaviors for different kinds of substrates and AFM tips and more accurate research is needed. In particular, the delicate role of non volatile material in water condensates should be carefully investigated.

\subsection{Measurements in sharp cracks}
\label{sec:MeasureForceCracks}

\begin{figure}[!h]
\centering
\includegraphics[width=0.45 \textwidth]{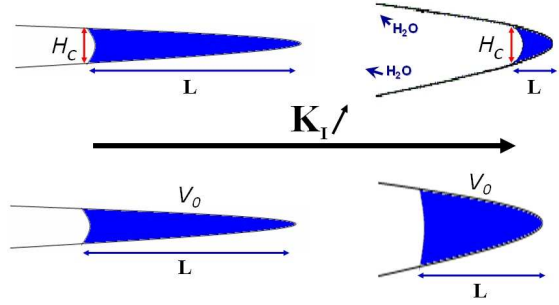}
\caption{Sketch of the variation of the condensation length after an increase of the stress intensity factor $K_I$: (top) case of equilibrium; (bottom) constant volume model. NB: exaggerated vertical scale.}
\label{fig:KLVarModels}
\end{figure}

Sharp crack tips in very brittle materials such as glasses have recently been proven to constitute an invaluable tool to study capillary condensation (Grimaldi \etal 2008). By using a stable testing configuration such as the DCDC specimen it is possible to obtain extremely flat crack surfaces (with roughness of the order of 0.4 nm RMS on a 1x1 $\mu$m$^2$ region) and to induce a controlled crack opening profile by carefully loading the sample in pure mode I (Pallares \etal 2009). We should highlight that although a nanometric roughness could be relevant affecting the capillary forces (cf.\ section \ref{sec:friction}), the complementarity of the roughness of brittle crack surfaces makes the local crack opening variations negligible. The crack opening $2u$ can thus be considered as a monotonously increasing function of the distance $X$ from the crack tip. In a close neighborhood of the crack tip this can be expressed by a parabolic shape (cf.\ Fig.\ \ref{fig:KLVarModels}) according to the Irwin equation:

\begin{equation}
2u(X,K_I) = {2K_I \over E'} \left( {8 X \over \pi} \right)^{1/2}
\label{eq:Irwin}
\end{equation}
where $K_I$ is the stress intensity factor and $E'$ an effective Young modulus.
The confinement in the sharp crack tips in glass is very elevated. Typical crack tip radii are of the order of 0.5 nm and the crack opening is of the order of 10 nm at 1 $\mu$m distance from the crack tip. This can be interpreted locally as a plane pore with progressive opening and it is thus expected to be filled with equilibrium capillary condensation up to a critical opening $H_c = D_c$ ($H_c=2r_k \cos \theta$ for partial wetting and $H_c=2 r_k + 3e$ for perfect wetting under the assumption that the liquid-solid interactions are of the vdW type, cf.\ insert D).

\begin{figure}[!h]
\centering
\begin{minipage}[t]{0.48\linewidth}
   \centering
   \includegraphics[width=8 cm]{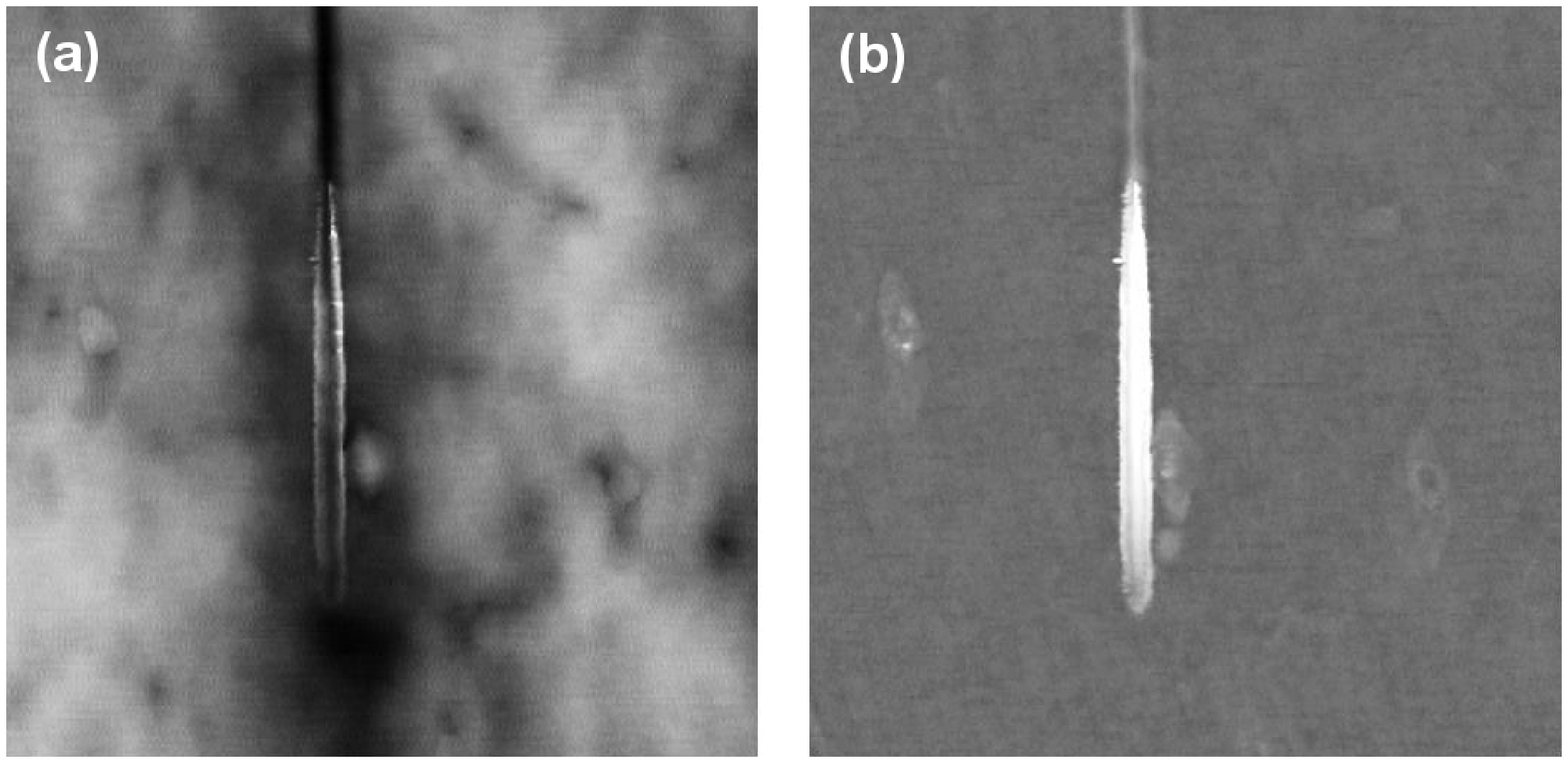}
   \caption{Typical AFM height (a) and phase (b) images of the crack tip for a fracture propagating from top to bottom of the image (1x1 $\mu$m). The linear gray scale range is respectively 10 nm and 10$^\circ$.}
   \label{fig:HeightPhaseCond}
\end{minipage}%
\hfill
\begin{minipage}[t]{0.48\linewidth}
   \centering
   \includegraphics[width=8 cm]{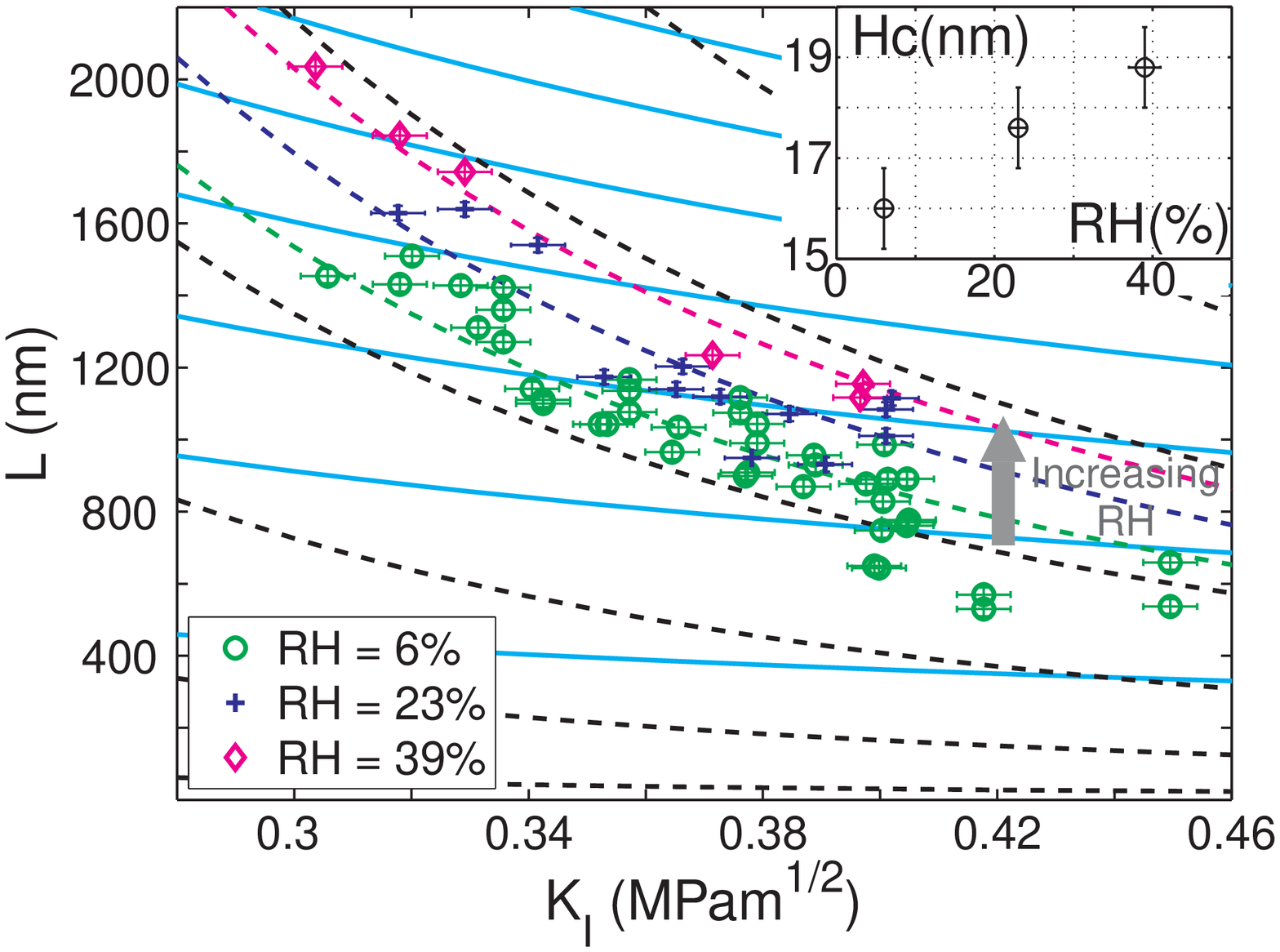}
   \caption{Plot of the condensation length $L$ versus $K_I$ for three different relative humidities (RH). Dashed curves represent the equilibrium model which is consistent with the data. Continuous curves represent the rejected constant volume model. The inset contains the fitted values of the critical distance $H_c$ for each relative humidity (after Grimaldi \etal 2008).}
   \label{fig:KLCurve}
\end{minipage}
\end{figure}

Since the critical distance $H_c$ is generally in the nanometer range, the length $L$ of the condensed phase is generally in the order of hundreds to thousands of nm and it has the advantage of significantly amplifying the effect of variations of $H_c$. Accurate measurements of $H_c$ can be obtained by measuring the variations of the condensation length that are induced by a modulation of the crack opening profile through changes of the sample loading as shown in Figs.\ \ref{fig:KLVarModels} and \ref{fig:KLCurve}. Such a measurement was recently made possible by the development of AFM phase imaging techniques (cf.\ insert G) that can detect the presence of capillary condensation in the crack cavity by measuring the energy dissipated in the formation of a liquid bridge between the condensed phase and the AFM tip during the imaging in tapping mode (Fig.\ \ref{fig:HeightPhaseCond}).

\singlespacing

%%%%%%%% Insert G
        \fbox{
\parbox{14.8cm}{
\begin{center}
\subsubsection*{Insert G: Capillary forces in AFM phase imaging (Tapping mode).}
\end{center}

High amplitude oscillating AFM techniques (tapping mode) are widely used in surface imaging due to high lateral resolution and reduced tip and substrate damage. The cantilever oscillation is stimulated by a high frequency piezo driver at a constant frequency $\omega$ near the cantilever resonant frequency $\omega_0$ ($f_0 = \omega_0/2\pi \simeq 330$ kHz) in order to have a free oscillation amplitude $A_0 \simeq 30$ nm at resonance.
During the scan of a surface, a feedback loop acts on the vertical position of the cantilever in order to maintain the amplitude of the tip oscillation constant to a set point value $A_{sp} \simeq 20$ nm. The vertical position of the cantilever motion provides the vertical topographic signal. Although the tip-sample interaction is non-linear, the motion of the tip can be well approximated by an harmonic oscillation $z(t) = A cos (\omega t + \theta)$. The phase delay $\theta$ between the stimulation and the oscillation of the cantilever can be related to the energy $E_{diss}$ which is dissipated during contact of tip and sample (Cleveland \etal 1998):

\begin{equation}
\sin \theta = {\omega \over \omega_0} {A_{sp} \over A_0} + {Q E_{diss} \over \pi k A_0 A_{sp}}
\label{eq:EnDiss}
\end{equation}
where $Q \sim 600$ and $k \sim 40$ N/m are the quality factor and the stiffness, respectively, of typical tapping cantilevers.
Phase imaging can thus provide a map of the local energy dissipation caused by the tip-sample interaction during the cantilever oscillation. This quantity contains several contributions related to viscoelastic or viscoplastic sample deformation and also from the local adhesion properties. However, for very stiff substrates such as glass or silicon, the bulk dissipation will be negligible and phase imaging will mostly be sensitive to adhesion and in particular to the formation and rupture of capillary bridges between the tip and the sample at each oscillation. Zitzler \etal (2002) have developed a model to relate the energy dissipation to the hysteretic behavior of the capillary force-distance curve. In particular, it was shown that the formation of capillary bridges has a very sensitive influence on the position of the transition between attractive and repulsive modes in the tapping regime (Ciccotti \etal 2008), and that this effect can be used to increase the sensitivity in the measurement of the local wetting properties.

On the other hand, capillary interactions can become a significant source of perturbation and artifacts in the topographical AFM images in tapping mode. Changes in the relative humidity during imaging can alter the stability of the mode of operation and induce an intermittent mode change which is the source of intense noise and local artifacts. In particular, in the presence of local patches with different wetting properties, the oscillation mode can systematically change between different areas of the image, inducing fake topographic mismatches. However, the topographical artifacts are of the order of the nanometer and are only relevant in very sensitive measurements on flat substrates. In such cases, an attentive monitoring of the phase imaging is required for detecting and avoiding the mode changes.
Another drawback of the formation of capillary bridges is the significant loss of lateral resolution caused by the increase of the interaction area between tip and sample (cf.\ the smearing of the crack-tip condensate in Fig.\ \ref{fig:HeightPhaseCond}).
Again, a careful tailoring of the imaging conditions can lead the repulsive interaction to be dominant and thus allows recovering a higher resolution.
}}
%%%%%%%% FIN Insert G
%\doublespacing

\newpage
\clearpage

The strongly negative capillary pressure in the crack cavity will apply important internal forces on the crack walls which add to the effect of the external forces and tend to close the crack. However, the advantage of such a stiff configuration is that even huge capillary pressures will have little effect on the crack opening profile when applied over a length $L$ of micrometer size near the crack tip. The above cited technique can thus be comfortably applied to first order determinations of $H_c$ with 10\% accuracy by neglecting the action of capillary forces. On the other hand, these forces can become more relevant if the external load (and thus the crack opening) are reduced enough to let the condensation extend to several tens of microns. In this context the deviations of the measured condensation length from the above cited prediction may be used to infer the value of the capillary pressure.

When implemented in controlled atmosphere, this technique can be used to measure both the critical condensation distance and the capillary pressure as a function of relative humidity and thus provide important information on the surface tension of the liquid, and on its Kelvin radius. This technique provides interesting complement to the SFA measurements since it does not suffer from the limitations of the SFA equipment to sustain the very strong negative pressures generated by the capillary meniscus at low humidity.

We remark that the values of the measured critical distance $H_c$ for water condensation in silica glass in Fig.\ \ref{fig:KLCurve} are quite large in relation to the prediction of eq.\ (\ref{eq:3e}), which should not exceed 3 nm at 70\% RH. This phenomenon of enhanced condensation is analog to  other similar observations, such as the measurement of large meniscus radii in SFA measurements between mica sheets (Christenson 1985) where capillary bridges of water condensed from a non polar liquid were argued to be enriched by ionic solute. These interesting effects still need more investigation and will provide important information on the chemical effects due to glass water interactions under tensile stress (cf.\ section \ref{sec:SurfaceChemistry}).

\section{Influence on Friction forces}
\label{sec:friction}

\subsection{Static friction and powder cohesion}
\label{sec:powders}

It is an everyday experience that a small amount of liquid can change drastically the physical properties of granular matter. Such effects are important not only on the beach for building sandcastles, but also in a number of industries as diverse as pharmaceutical, construction and agriculture, and in a number of geophysical applications.
The changes induced by a small amount of liquid in granular materials are primarily caused by adhesive forces associated with interstitial liquid bridges between grains (Zimon 1967, Bocquet \etal 2002).
In the absence of cohesion forces the stability of a heap of beads can be described by a modified Mohr-Coulomb analysis, thus predicting a maximum angle of stability $\theta_m=\phi$, where $\mu = tan \phi = \tau/\sigma$ is the static friction coefficient between the beads. The effect of an adhesive stress $c$ between the beads can be modeled by writing a modified Coulomb yield criterion $\tau = \mu(\sigma + c)$ and thus obtaining an implicit equation for the critical angle $\theta_m$ (Halsey and Levine 1998):

\begin{equation}
\tan \theta_m = \tan \phi \left( 1 + {c \over \rho g h \cos \theta_m} \right)
\label{eq:CritAngle}
\end{equation}
where $\rho$ and $h$ are the density and height of the granular layer.
The critical angle is thus an increasing function of the adhesive stress $c$ and it saturates to $\pi/2$ (stuck heap) when $\mu c > \rho g h$.

Several experiments have recently been investigating the applicability of this relation and the consistency of the model with capillarity based adhesion forces. A first series of experiments has shown that the injection of a very small fixed amount of non volatile liquids to a granular heap can significantly increase the critical angle of stability (Hornbaker \etal 1997, Halsey and Levine 1998). A different kind of experiments has then investigated the more subtle effect of interparticle liquid bridges formed by condensation from a humid atmosphere (Bocquet \etal 1998, Fraysse \etal 1999).
While the first kind of experiments can not be related to equilibrium quantities, the second kind has allowed a deep investigation of the first order transition of capillary condensation.

The critical angle was shown to be logarithmically increasing function of the aging time $t_w$ (after shaking) of a granular heap (Bocquet \etal 1998):

\begin{equation}
\tan \theta_m = \tan \theta_0 + \alpha {\log_{10} t_w \over \cos \theta_m}
\label{eq:CritAngleAging}
\end{equation}

The coefficient $\alpha$ is a measure of the aging behavior of friction in the granular medium and it was shown to be an increasing function of relative humidity, being substantially null in dry air. Moreover, the aging behavior was shown to be enhanced by both increasing the rest angle of the heap during the aging period and by intentionally wearing the particles (by energetic shaking) before the measurements (Restagno \etal 2002).

The logarithmic time dependency of the adhesion forces was modeled as an effect of the dynamic evolution of the total amount of condensed water related to progressive filling of the gaps induced by the particle surface roughness (cf.\ insert H).
The increase of the aging rate $\alpha$ with the rest angle before the measurements can be explained by the effect of a series of small precursor beads adjustments, inducing a modification of the contacts and of the condensed bridges population. The effect of wear of the particles can be explained by accounting for both the increased roughness of the beads and the presence of a wear induced dust consisting in small particles that significantly enhance the nucleation of further bridges (Restagno \etal 2002).

%\bigskip

\subsection{Time and velocity dependence in nanoscale friction forces}

Sliding friction is an everyday life issue, and its universal nature emerges from the great variety of industrial processes and natural phenomena in which it plays a central role (Persson 2000). With the miniaturization of moving components in many technological devices, such as microelectromechanical systems and hard disks, it has become of primary importance to study surface forces like friction, viscous drag and adhesion at microcales and nanoscales. The relevance of surface forces is greatly enhanced with regard to volume forces when the spatial scale is reduced to nanometers, but another major physical change comes from the increasing role of thermal fluctuations in the surface processes.

AFM has become the most efficient tool to study surface forces at the nanoscale, and the AFM tip sliding on a surface can often be considered as a model system for technologically relevant devices. The terminal apex of typical AFM tips can be roughly approximated by a sphere of radius between 10 and 100 nm, and the AFM contact imaging of a nanoscale rough surface can provide both a measure of its roughness through the vertical deflection of the probe cantilever and a measure of the friction forces through the dependence of the lateral deflection of the cantilever on the scan speed and on the normal applied load.

Nanoscale friction was rapidly shown not to respect the Amonton laws, being dependent on both sliding velocity and normal load. Moreover, it was shown to be strongly affected by the nanoscale roughness of the substrate and by the wetting properties of both the AFM tip and the substrate as well as by   relative humidity. Nanofriction measurements have long been controversial, but recent careful measurements in controlled atmosphere have allowed defining a clearer scenario.

The sliding kinetics of an AFM tip has been shown to be determined by both the thermally activated stick-slip dynamics of the AFM tip on the substrate and the time dependent formation of capillary bridges between the tip and the asperities of the rough substrate (cf.\ insert H).

\singlespacing

%%%%%%%% Insert H
        \fbox{
\parbox{14.8cm}{
\begin{center}
\subsubsection*{H. Effect of roughness on adhesion forces.}
\end{center}

The first order effect of roughness is to screen the interactions between surfaces, with an increased efficiency for the shorter range interaction. The molecular range solid-solid interactions are thus very efficiently screened by nanoscale roughness, while the capillary interactions have a more subtle behavior. Three main regimes were identified by Halsey and Levine (1998) as a function of the volume of liquid $V$ available for the formation of a capillary bridge between two spheres: (1) the {\it asperity regime} prevails for small volumes, where the capillary force is dominated by the condensation around a single or a small number of asperities; (2) the {\it roughness regime} governs the intermediate volume range, where capillary bridges are progressively formed between a larger number of asperities and the capillary force grows linearly with $V$ (as in Hornbaker \etal 1997); (3) the {\it spherical regime} where eq.\ (\ref{eq:pull-off-partial}) for the force caused by a single larger meniscus is recovered.
The extension of the three domains is determined by the ratio between the characteristic scales of the gap distribution (height $l_R$ and correlation length $\xi$) and the sphere radius $R$. When dealing with capillary bridges in equilibrium with undersaturated humidity, the roughness regime should be reduced to a narrow range of humidity values such that $l_R \sim r_k$, in which the capillary force jumps from a weak value to the spherical regime value (as in the experiments of Fraysse \etal 1999).

However, due to the first order nature of capillary condensation transition, the equilibrium condition may be preceded by a long time dependent region where the capillary bridges between asperities are progressively formed due to thermal activation. The energy barrier for the formation of a liquid bridge of volume $v_d \sim h A$ between asperities of curvature radius $R_c$ may be expressed as (Restagno \etal 2000):

\parbox{6cm}{
$$\Delta \Omega = v_d \rho_l \Delta \mu = v_d \rho_l k_B T \log{P_{sat} \over P_v}$$
The probability that condensation occurs before a time $t_w$ is:
$$\Pi(t_w) = 1 - \exp\left(-{t_w \over \tau}\right)$$
$$\tau = \tau_0 \exp\left( \Delta\Omega \over k_B T \right)$$}
\parbox{1cm}{}
\parbox{6cm}{\includegraphics[width=0.45 \textwidth]{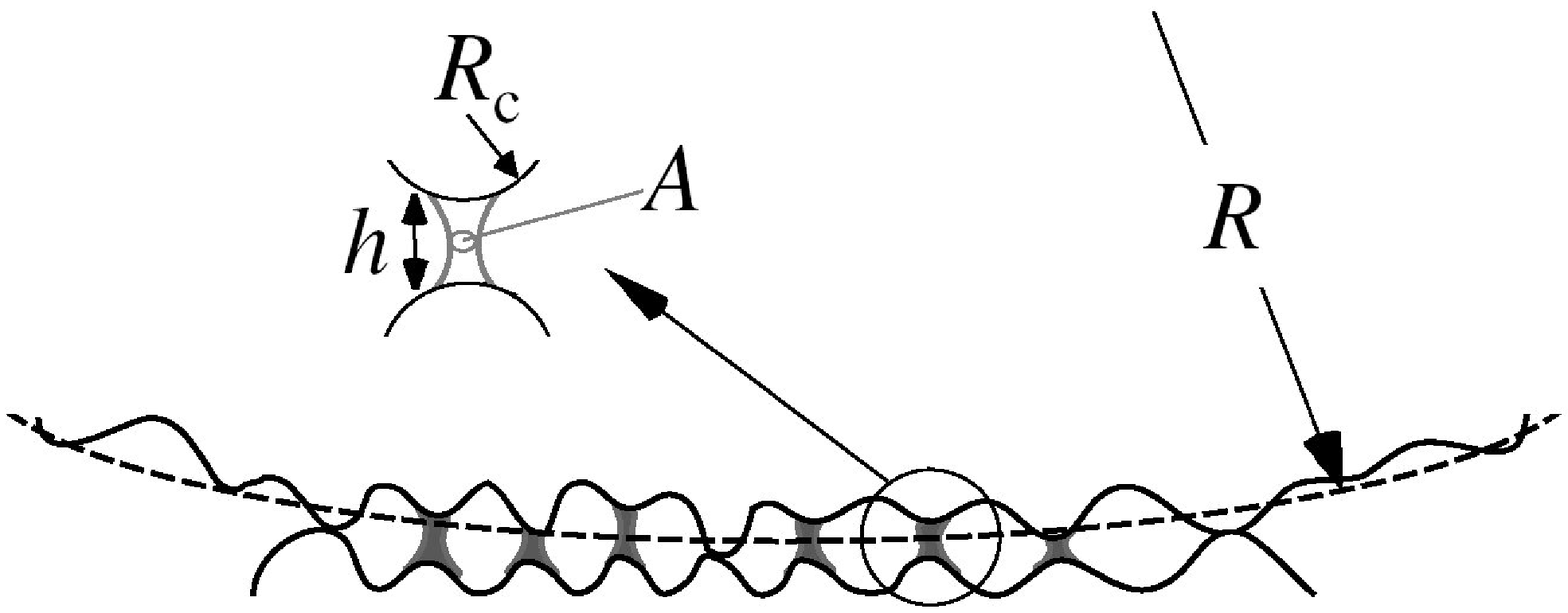}}

By integrating over a roughness dependent distribution of $N_{tot}$ nucleation sites, each one contributing with a force $2 \pi \gamma R_C$, an expression for the total force can be derived (Bocquet \etal 2002) that predicts a logarithmic increase as a function of the aging time $t_w$:

\begin{equation}
F_C = F_0 + 2 \pi \gamma R_c N_{tot} {k_B T \over V_0 \rho_l \Delta \mu} \log {t_w \over \tau_0}
\label{eq:FCapDepT}
\end{equation}
where $V_0$ is a roughness dependent range for the distribution of the individual liquid bridge volume $v_d$.

}}
%%%%%%%% FIN Insert H
%\doublespacing

%\bigskip

When relative humidity is low, or the substrate is weakly wettable, stick-slip sliding has a dominant effect and results in a positive logarithmic dependence of the friction force on the sliding velocity (Gnecco \etal 2000, Riedo \etal 2003).
The AFM tips keeps being stuck on nanoscale asperities and intermittently slips when thermal fluctuations allow skipping a local energy barrier which is progressively reduced by the accumulation of elastic energy due to the scanning. The dynamic friction force can be described by the following equation:

\begin{equation}
F_F = \mu(F_N + F_{SS}) + \mu[F_c(t)] + m \log \left( v \over v_B \right)
\label{eq:Frict1}
\end{equation}
where $F_N$ is the normal force, $F_{SS}$ the solid-solid adhesion force within the liquid, $F_c(t)$ describes the eventual presence of time dependent capillary forces and the last term describes the positive velocity dependence induced by stick-slip motion.

For more hydrophilic substrates, higher relative humidity, or increasing substrate roughness, the formation of capillary bridges and wetting films deeply modifies the friction dynamics letting the friction force be a logarithmically decreasing function of the sliding velocity (Riedo \etal 2002). This effect was successfully explained by applying the modeling developed by Bocquet \etal (1988) to account for the time dependent thermally activated formation of capillary bridges between the nanoscale asperities of both the probe and the substrate.
When the two rough surfaces are in relative sliding motion, the proximity time $t_w$ of opposing tip and substrate asperities is a decreasing function of the sliding velocity $v$. The number of condensed capillary bridges (and thus the total capillary force) is thus expected to be a decreasing function of the sliding velocity according to eq.\ (\ref{eq:FCapDepT}), and this trend should be reflected in the dynamic friction force according to eq.\ (\ref{eq:Frict1}).

The study of AFM sliding friction forces has thus become an important complementary tool to study the time and load dependence of capillary forces. Notably, the friction forces were shown to present a 2/3 power law dependence on the applied load $F_N$ (Riedo \etal 2004) and an inverse dependence on the Young modulus $E$ of the substrate (Riedo and Brune, 2003). These two effects were both explained by the increase of the nominal contact area where capillary bridges are susceptible to be formed, when either the normal load is increased, or the Young modulus of the substrate is decreased. The following equation for the capillary force during sliding was proposed in order to account for all these effects:

\begin{equation}
F_c = 8 \pi \gamma_{lv} R (1 + KF_N^{2/3}) \left( 1 \over \rho_l l_R R_c^2 \right)
{ \log{v_0 \over v} \over \log{P_{sat} \over P_v}} \qquad K = {1 \over r_K} \left( {9 \over 16 R E^2}  \right)^{1/3}
\label{eq:Frict2}
\end{equation}
the variables being defined as in the insert H.
Quantitative measurements of the AFM friction forces were thus shown to be useful in determining several physical parameters of interest, such as an estimation of the AFM tip radius and contact angle or important information on the activation energy for the capillary bridge formation (Szoskiewicz and Riedo 2005).

The formation of capillary bridges can have significant effects on the AFM imaging in contact mode due to the variations of the contact forces and consequently of the lateral forces during the scan. Thundat \etal (1993) have investigated the effect of humidity on the contrast when measuring the atomic level topography of a mica layer. The topographic contrast is shown to decrease with humidity above 20\% RH due to an increase of the lateral force that acts in deforming the AFM cantilever and thus influences the measurement of the vertical deflection.

\bigskip

\subsection{Friction forces at macro scale}

Capillary forces can also affect the friction properties between macroscopic objects. In dry solid friction, ageing properties have been studied on various materials, and have been related to the slow viscoplastic increase of the area of contact between asperities induced by the high values of the stress in the contact region (Baumberger \etal 1999). However, the importance of humidity has been reported by geophysicists in rock onto rock solid friction (Dieterich and Conrad, 1984).

In the presence of a vapor atmosphere, the static friction coefficient was shown to increase logarithmically with the contact time, while the dynamic friction coefficient was shown do decrease with the logarithm of the sliding velocity (Dieterich and Conrad, 1984). The first effect is analog to the aging behavior of the maximum contact angle in granular matter as discussed in section \ref{sec:powders}. The second effect is analog to what observed in the sliding friction of a nanoscale contact as discussed in the previous section. However, the general behavior is strongly modified due to the greater importance of the normal stresses that induce significant plastic deformation at the contact points. The aging behavior must then be explained by the combined action of the evolution of the contact population due to plastic deformation and the evolution of the number of capillary bridges due to thermally activated condensation. This induces a more complex dependence on the normal load, since this modifies both the elastic contact area and the progressive plastic deformation of the contacts, and thus influences the residual distribution of the intersurface distances that govern the kinetics of capillary condensation.

Based on these experimental observations, Rice and Ruina (1983) have proposed a phenomenological model for non stationary friction, in which friction forces depend on both the instantaneous sliding velocity $v$ and a state variable $\varphi$ according to:

\begin{equation}
F(v,\varphi) = F_N \left[ \mu_0 + A \log \left( v \over v_0 \right) + B \log \left( \varphi v_0 \over d_0 \right) \right]
\label{eq:RateState}
\end{equation}

\begin{equation}
\dot{\varphi} = 1 - {v \varphi \over d_0}
\label{eq:RateState2}
\end{equation}
where $A$ and $B$ are positive constants, while $d_0$ and $v_0$ are characteristic values of the sliding distance and velocity. The first term accounts for the logarithmic dependence on the sliding velocity, while the second term accounts for the logarithmic dependence on the static contact time through the evolution of the state variable $\varphi$ according to eq.\ (\ref{eq:RateState2}). This phenomenological modeling can be applied to other intermittent sliding phenomena (peeling of adhesives, shear of a granular layer, etc.) and the significance of the state variable $\varphi$ is not determined a priori. However, in the case of solid friction, the state variable $\varphi$ can be related to the population of microcontacts and capillary bridges (Dieterich and Kilgore, 1994).

%\newpage

\section{Influence on Surface Chemistry}
\label{sec:SurfaceChemistry}

The presence of nanometric water films on solid surfaces has deep consequences on the surface physical and chemical properties. Chemistry in these extremely confined layers is quite different than in bulk liquids due to the strong interaction with the solid surface, to the presence of the negative capillary pressure, to the reduction of transport coefficients and to the relevance of the discrete molecular structure and mobility that can hardly be represented by a continuum description. The role of thermal fluctuations and their correlation length also become more relevant.

Thin water films can have a major role in the alteration of some surface layer in the solid due to their effect on the local dissolution, hydration, oxidation, hydrolysis, and lixiviation which are some of the basic mechanisms of the corrosion processes.
Water condensation from a moist atmosphere is quite pure and it is thus initially extremely reactive towards the solid surface. However, the extreme confinement prevents the dilution of the corrosion products, leading to a rapid change in the composition and pH of the liquid film. Depending on the specific conditions this can either accelerate the reaction rates due to increased reactivity and catalytic effects, or decelerate the reaction rate due to rapid saturation of the corrosion products in the film. This condition of equilibrium between the reactions of corrosion and recondensation can  lead to a progressive reorganization of the structure of the surface layer in the solid. The extreme confinement and the significance of the fluctuations can cause the generation of complex patterns related to the dissolution-recondensation process, involving inhomogeneous redeposition of different amorphous, gel or crystalline phases (Watanabe \etal 1994, Christenson and Israelachvili 1987).

\begin{figure}[!h]
\centering
\includegraphics[width=0.45 \textwidth]{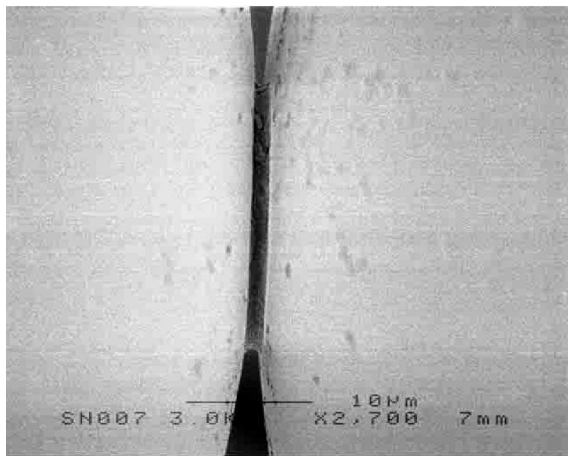}
\caption{MEB photograph of a solid bridge between two glass beads (magnification x2700, from Olivi-Tran \etal 2002).}
\label{fig:SolidBridge}
\end{figure}

The dissolution-recondensation phenomenon also happens at the capillary bridges between contacting solid grains or between the contacting asperities of two rough solid surfaces. When humidity undergoes typical ambient oscillations, capillary bridges and films are formed or swollen in moist periods, inducing an activity of differential dissolution. The subsequent redeposition under evaporation in more dry conditions is particularly effective in the more confined regions, i.e.\ at the borders of the solid contact areas, acting as a weld solid bridge between the contacting parts (cf.\ Fig.\ \ref{fig:SolidBridge}). This can be responsible of a progressive increase in the cohesion of granular matters, which has important applications in the pharmaceutical and food industry, and of the progressive increase of the static friction coefficient between contacting rocks.

Another domain where the formation of capillary condensation has a determinant impact on the mechanical properties is the stress-corrosion crack propagation in moist atmosphere (cf.\ Ciccotti 2009 for a review). We already mentioned in section \ref{sec:MeasureForceCracks} that the crack tip cavity in brittle materials like glass is so confined that significant capillary condensation can be observed at its interior. During slow subcritical crack propagation, the crack advances due to stress-enhanced chemical reactions of hydrolysation and leaching that are deeply affected by the local crack tip environment. Capillary condensation has a fundamental impact on several levels on the kinetics of this reaction: (1) the presence of a liquid phase makes the preadsoprion of water molecules near the crack tip easier; (2) the negative Laplace pressure determines the chemical activity of the water molecules in the meniscus and directly affects the reaction rate; (3) the confined nature of the condensation along with its limited volume are responsible of an evolution of the chemical composition of the condensate that has a direct and major effect on the corrosion reactions at the crack tip, especially by changes of the pH and by the enrichment in alkali species due to stress-enhanced leaching (Célarié \etal 2007).

\section*{References}

%\begin{thebibliography}{1}

%\begin{harvard}

Abramowitz, M. and Stegun, I. A., eds. 1972. {\it Handbook of Mathematical Functions with Formulas, Graphs, and Mathematical Tables}. New York: United States Government Printing.

Barrett, E. P., Joyner, L. G. and Halenda, P. P. 1951. The determination of pore volume and area distributions in porous substances. 1. computation from nitrogen isotherms. {\it J. Am. Chem. Soc.} 73: 373--80.

Baumberger, T., Berthoud, P. and Caroli C. 1999. Physical analysis of the state- and rate-dependent friction law. II. Dynamic friction. {\it Phys. Rev. B} 60: 3928--39.

Bhushan, B. (ed), 2007. {\it Springer Handbook of Nanotechnology}. New York: Springer. 2nd ed.

Bocquet, L., Charlaix, E., Ciliberto, S. and Crassous J. 1998. Moisture-induced ageing in granular media and the kinetics of capillary condensation. {\it Nature} 296: 735--7.

Bocquet, L., Charlaix, E. and Restagno, F. 2002 Physics of humid granular media. {\it C. R. Physique} 3: 207--15.

Bowden, F. P., and Tabor, D., 1950. {\it Friction and lubrication in solids}. Oxford: Clarendon Press.

Brunauer S., Emmett, P. H. and Teller, E. 1938, Adsorption of gases in multimolecular layers. {\it J. Am. Chem. Soc.} 60: 309--19.

Célarié, F., Ciccotti, M. and Marlière, C. 2007. Stress-enhanced ion diffusion at the vicinity of a crack tip as evidenced by atomic force microscopy in silicate glasses. {\it J. Non-Crist. Solids.} 353: 51-68.

Charlaix, E. and Crassous, J. 2005. Adhesion forces between wetted solid surfaces. {\it J. Chem. Phys} 122: 184701.

Christenson, K. 1988. Adhesion between surfaces in undersaturated vapors--A reexamination of the influence of meniscus curvature and surface forces. {\it J. Coll. Interf. Science} 121: 170--8.

Christenson, H. K., and Israelachvili, J. N. 1987. Growth of ionic crystallites on exposed surfaces. {\it J. Coll. Interf. Science} 117: 576--7.

Christenson, H. K. 1985. Capillary condensation in systems of immiscible liquids. {\it J. Coll. Interf. Science} 104: 234--49.

Christenson, H. K. 1994. Capillary condensation due to van der Waals attraction in wet slits. {\it Phys. Rev. Lett.} 73: 1821--4.

Ciccotti, M., George, M., Ranieri, V., Wondraczek, L. and Marlière, C. 2008. Dynamic condensation of water at crack tips in fused silica glass.
{\it J.\ Non-Cryst.\ Solids.} 354: 564--8.

Ciccotti, M. 2009. Stress-corrosion mechanisms in silicate glasses. {\it J. Phys. D: Appl. Phys.} In press. \verb|arXiv:0901.2809|.

Cleveland, J. P., Anczykowski, B., Schmid, A. E. and Elings, V. B. 1998. Energy dissipation in tapping-mode atomic force microscopy. {\it Appl.\ Phys.\ Lett.} 72: 2613--5.

Crassous, J. 1995. PhD Thesis. Ecole Normale Supérieure de Lyon, France.

Crassous, J., Charlaix, E. and Loubet J. L. 1994. Capillary condensation between high-energy surfaces: experimental study with a surface force apparatus. {\it Europhys. Lett.} 28: 37--42.

de Gennes, P. G. 1985. Wetting: statics and dynamics. {\it Rev. Modern Phys.} 57: 827--63.

de Gennes, P. G., Brochard, F. and Quere, D. 2003. {\it Capillarity and Wetting Phenomena: drops, bubbles, pearls, waves}. New York: Springer.

de Lazzer, A., Dreyer, M. and Rath, H. J. 1999. Particle-surface capillary forces. {\it Langmuir} 15: 4551--9.

Derjaguin, B. V. 1955. {\it Kolloidn. Zh.} 17: 191.

Derjaguin, B. V. and Churaev, N. V. 1976. Polymolecular adsorption and capillary condensation in narrow slit pores
{\it J. Coll. Interf. Science} 54: 157--75.

Dieterich, J. H. and Conrad, G. 1984. Effect of humidity on time-dependent and velocity-dependent friction in rocks. {\it J. Geophys. Res.} 89: 4196--202.

Dieterich, J. H. and Kilgore, B. D. 1994. Direct observation of frictional contacts-new insights for state-dependent properties. {\it Pure Appl. Geophys.} 143: 283--302.

Fisher, L. R., Israelachvili, J. N. 1981. Direct measurement of the effect of meniscus force on adhesion: a study of the applicability of macroscopic thermodynamis to microscopic liquid interfaces. {\it Coll. Surf.} 3: 303--19.

Fraysse, N., Thomé, H. and Petit, L., 1999. Humidity effects on the stability of a sandpile. {\it Eur. Phys. J. B} 11: 615--9.

Halsey, T. C. and Levine, A. J. 1998. How sandcastles fall. {\it Phys. Rev. Lett.} 80: 3141--4.

Heilmann, R., Ahn, M. and Schattenburg, H., 2008. Nanomirror array for high-efficiency soft x-ray spectroscopy.
SPIE Newsroom. 27 August 2008. DOI: \verb|10.1117/2.1200808.1235|.

Hornbaker, D. R., Albert, I., Barabasi, A. L. and Shiffer, P. 1997 What keeps sandcastles standing? {\it Nature} 387: 765--6.

Gelb, L. D., Gubbins, K. E., Radhakrishnan, R. and Sliwinska-Bartkowiak, M. 1999. Phase separation in confined systems. {\it Rep. Progr. Phys.} 62: 1573--659.

Grimaldi, A., George, M., Pallares, G., Marlière, C. and Ciccotti, M. 2008. The crack tip: a nanolab for studying confined liquids. {\it Phys. Rev. Lett.} 100: 165505.

Israelachvili, J. N. 1992. {\it Intermolecular and surface forces}. New York: Academic Press. 2nd ed.

Kohonen, M. M., Maeda, N. and Christenson, H. K. 1999. Kinetics of capillary condensation in a nanoscale pore. {\it Phys. Rev. Lett.} 82: 4667--70.

Lefevre, B., Sauger, A., Barrat, J. L., Bocquet, L., Charlaix, E., Gobin, P. F. and Vigier, G. 2004. Intrusion and extrusion of water in hydrophobic mesopores. {\it J. Chem. Phys.} 120: 4927--38.

Maugis, D. 1992. Adhesion of spheres: the JKR-DMT transition using a dugdale model. {\it J. Coll. Interf. Science} 150: 243--269.

Maugis, D. and Gauthier-Manuel, B. 1994. JKR-DMT transition in the presence of a liquid meniscus. {\it J. Adhes. Sci. Technol.} 8: 1311--22.

Orr, F. M., Scriven, L. E., Rivas, A. P. 1975. Pendular rings between solids-Menuscus properties and capillary force. {\it J. Fluid Mech.} 67: 723--42.

Pakarinen, O. H., Foster, A. S., Paajanen, M., Kalinainen, T., Katainen, J., Makkonen, I., Lahtinen, J. and Nieminen, R. M. 2005. Towards an accurate description of the capillary force in nanoparticle-surface interactions {\it Modelling Simul. Mater. Sci. Eng.} 13: 1175--1186.

Pallares, G., Ponson, L., Grimaldi, A., George, M., Prevot, G. and Ciccotti, M., 2009. Crack opening profile in DCDC specimen. {\it Int. J. Fract.}. In press. \verb|arXiv:0903.5192|.

Restagno, F., Bocquet, L. and Biben, T. 2000. Metastability and nucleation in capillary condensation. {\it Phys. Rev. Lett.} 84: 2433--6.

Restagno, F., Ursini, C., Gayvallet, H. and Charlaix, E. 2002. Aging in humid granular media. {\it Phys. Rev. E} 66: 021304.

Rice J. R. and Ruina A. L. 1983. Stability of steady frictional slipping. {\it J. Appl. Mech.} 50: 343--9.
%JOURNAL OF APPLIED MECHANICS-TRANSACTIONS OF THE ASME   Volume: 50   Issue: 2   Pages: 343-349   Published: 1983

Riedo, E., Lévy, F. and Brune, H., 2002. Kinetics of capillary condensation in nanoscopic sliding friction. {\it Phys. Rev. Lett.} 88: 185505.

Riedo, E., Gnecco, E., Bennewitz, R., Meyer, E. and Brune, H. 2003. Interaction potential and hopping dynamics governing sliding friction. {\it Phys. Rev. Lett.} 91: 084502.

Riedo, E., Palaci, I., Boragno, C. and Brune, H. 2004. The 2/3 power law dependence of capillary force on normal load in nanoscopic friction. {\it J. Phys. Chem. B} 108: 5324--8.

Rowlinson, J. S. and Widom, B. 1982. {\it Molecular theory of capillarity}, Oxford: Clarendon Press.

Szoskiewicz, R. and Riedo, E. 2005. Nucleation time of nanoscale water bridges. {\it Phys. Rev. Lett.} 95: 135502.

Thomson, W. 1871. On the Equilibrium of Vapour at a Curved Surface of Liquid. {\it Phil. Mag.} 42: 448--52.

Thundat, T., Zheng, X. Y., Chen G. Y. and Warmack R. J. 1993. Role of relative humidity in atomic force microscopy imaging. {\it Surf. Sci. Lett.} 294: L939--943.

Torralvo, M. J., Grillet Y., Llewellyn P. L. and Rouquerol F. 1998. Microcalorimetric study of argon, nitrogen and carbon monoxyde adsorption on mesoporous vycor glass. {\it J. Coll. Interf. Science} 206: 527--31.

Verdaguer, A., Sacha, G. M., Bluhm, H. and Salmeron M. 2006. Molecular structure of water at interfaces: wetting at the nanometer scale.
{\it Chem. Rev.} 106: 1478--510.

Wan, K. T., Smith, D. T., Lawn, B. R. 1992. Fracture and contact adhesion energies of mica-mica, silica-silica, and mica-silica interfaces in dry and moist atmospheres. {\it J. Am. Ceram. Soc.} 75: 667--76.

Watanabe, Y., Nakamura, Y., Dickinson, J. T. and Langford, S. C. 1994. Changes in air exposed fracture surfaces of silicate glasses observed by atomic force microscopy. {\it J. Non-Cryst. Solids} 177: 9--25.

Xiao, X. and Qian, L. 2000. Investigation of humidity-dependent capillary force. {\it Langmuir} 16: 8153--8.

Xu, L., Lio, A., Hu, J., Ogletree, D. F. and Salmeron, M. 1998. Wetting and capillary phenomena of water on mica. {\it J. Phys. Chem. B} 102: 540--8.

Zimon, A. D. 1969. {\it Adhesion of dust and Powder}. New York: Plenum Press.

Zitzler, L., Herminghaus, S., and Mugele, F. 2002. Capillary forces in tapping mode atomic force microscopy. {\it Phys. Rev. B} 66: 155436.

%\end{thebibliography}
%\end{harvard}

\end{document}